\documentclass{article}

\usepackage{arxiv}

\usepackage[utf8]{inputenc} 
\usepackage[T1]{fontenc}    
\usepackage{url}            
\usepackage{booktabs}       
\usepackage{amsfonts}       
\usepackage{nicefrac}       
\usepackage{microtype}      
\usepackage{graphicx}
\graphicspath{{./figures/}}
\usepackage{bm}
\usepackage{amsmath, amssymb}
\usepackage{enumitem}
\usepackage{import}
\usepackage{subcaption}
\usepackage{xcolor}
\usepackage{appendix}

\newcommand{\xw}{\left({\bm x},{\bm \omega} \right)}

\title{Multi-view polarimetric scattering cloud tomography and retrieval of droplet size}

\author{Aviad Levis\thanks{Now at the Computing and Mathematical Sciences Department, California Institute of Technology, Pasadena, CA, 91125.}\\
	Viterbi Faculty of Electrical Engineering\\
	Technion - Israel Institute of Technology \\
	Haifa, 3200003, Israel\\
	{\tt\small aviad.levis@gmail.com}
	\And
	Yoav Y.~Schechner\\
	Viterbi Faculty of Electrical Engineering\\
	Technion - Israel Institute of Technology \\
	Haifa, 3200003, Israel\\
	{\tt\small yoav@ee.technion.ac.il}
	\And
	Anthony B.~Davis\\
	Jet Propulsion Laboratory\\
	California Institute of Technology\\
	Pasadena, CA, 91109\\
	{\small \tt Anthony.B.Davis@jpl.nasa.gov}
	\And
	Jesse Loveridge\\
	Dept. of Atmospheric Sciences\\
	University of Illinois\\
	Champaign, IL, 61820\\
	{\small \tt jesserl2@illinois.edu }
}

\begin{document}
	\maketitle
	
	\begin{abstract}
		Tomography aims to recover a three-dimensional (3D) density map of a medium or an object.
		In medical imaging, it is extensively used for diagnostics via X-ray computed tomography (CT).
		Optical diffusion tomography is an alternative to X-ray CT that uses multiply scattered light to deliver coarse density maps for soft tissues. We define and derive tomography of cloud droplet distributions via passive remote sensing. We use multi-view polarimetric images to fit a 3D polarized radiative transfer (RT) forward model. Our motivation is 3D volumetric probing of vertically-developed convectively-driven clouds that are ill-served by current methods in operational passive remote sensing.
		These techniques are based on strictly 1D RT modeling and applied to a single cloudy pixel, where cloud geometry is assumed to be that of a plane-parallel slab.
		Incident unpolarized sunlight, once scattered by cloud-droplets, changes its polarization state according to droplet size.
		Therefore, polarimetric measurements in the rainbow and glory angular regions can be used to infer the droplet size distribution.
		This work defines and derives a framework for a full 3D tomography of cloud droplets for both their mass concentration in space and their distribution across a range of sizes.
		This 3D retrieval of key microphysical properties is made tractable by our novel approach that involves a restructuring and differentiation of an open-source polarized 3D RT code to accommodate a special two-step optimization technique.
		Physically-realistic synthetic clouds are used to demonstrate the methodology with rigorous uncertainty quantification.
	\end{abstract}
	
	
	\section{Introduction \& Outline}
	\label{sec:introduction}
\begin{figure}[t]
		\centering
		\includegraphics[width=0.6\linewidth]{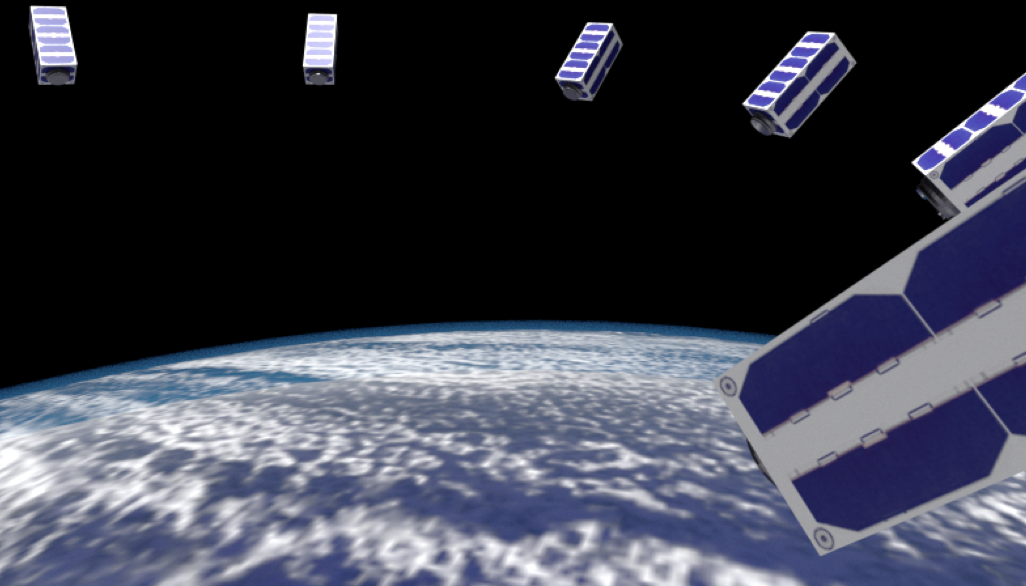}
		\caption{Artist’s illustration of the CloudCT~\cite{cloudct} mission: a distributed multi-view system of 10 nano-satellites orbiting the Earth in formation. Measurements acquired by the formation will enable tomographic retrievals of cloud properties.}
		\label{fig:cloudct}
	\end{figure}

	Clouds play a significant role at local and global scales, affecting weather, the water cycle, solar power generation, and impacting Earth's energy balance~\cite{trenberth2009earth}.
	Moreover, uncertainties in global climate models are significantly affected by our limited understanding, and therefore modeling, of cloud dynamics and microphysics~\cite{boucher2013clouds}.
	Thus, understanding, modeling, and predicting cloud properties is a key issue with worldwide socio-economic implications that is in the center of many research studies~\cite{rosenfeld1998satellite}.
	Much of the current understanding relies on routine remote sensing of cloud properties such as by the MODerate resolution Imaging Spectrometer (MODIS)~\cite{Platnick2003MODIS}.
	In practice, global-scale retrievals have so far been based on an individual pixel basis, using a crude approximation that clouds are plane-parallel slabs.
	This approximation uses a 1D radiative transfer (RT) model, which leads to biases in many retrievals~\cite{marshak2006impact} while other retrievals simply fail~\cite{cho2015frequency}.
	Convective clouds are therefore a blind spot due to their 3D nature.\footnote{
		Shallow convective clouds in the planetary boundary layer are also overlooked due to their unresolved scale in low-resolution sensors.}
	In its 2018 \emph{Decadal Strategy for Earth Observation from Space}~\cite{NAS_ESDS_2017}, the National Academies of Sciences, Engineering, and Medicine have indeed identified ``Clouds, Convection, and Precipitation'' as one of its five top-priority Targeted Variables for NASA's next generation of satellite missions.
	To bridge this gap, new technology is needed to study clouds as 3D volumetric objects, on a global scale. 
	The {\em CloudCT}~\cite{cloudct} space mission, by the European Research Council (ERC) is specifically destined to provide data and products for this goal. 
	It will involve 10 nano-satellites orbiting in formation, thus acquiring simultaneously unique multi-view measurements of such vertically-developed 3D clouds (Fig.~\ref{fig:cloudct}).
	
	Moreover, common retrieval of cloud droplet characteristics use two optical 
	bands simultaneously~\cite{nakajima1990determination}: a visible band, where reflected radiance increases with cloud optical thickness, and a shortwave IR (SWIR) band, where absorption by condensed water depends on cloud droplet size. To sense droplet size in 3D by CloudCT or other future missions, sensors will need to have either SWIR or polarization capability. 
	
	
	\subsection{Why polarized light?}
	
	There is an additional caveat in common retrievals, which rely on SWIR absorption~\cite{nakajima1990determination}. In addition to absorption, light undergoes multiple scattering in clouds. Multiple scattering diminishes sensitivity to droplet microphysics. High sensitivity to microphysics is embedded in single-scattering events. It is thus beneficial to pick-up single-scatter signals, out of the strong multiply-scattered background radiance. Polarization signals of scattered light are dominated by single-scattering events, and are thus highly sensitive to the type and size specifications of scatters. Thus in recent years, there is growing interest in polarimetric imagers for remote sensing of clouds and aerosols~\cite{deschamps1994polder, BreonGoloub_98, kalashnikova2011sensitivity, lukashin2013uncertainty, diner2013airborne, diner2018advances}. 
	In turn, increased interest in polarimetric sensing capabilities has led to the development of 1D and 3D polarized (or ``vector'') RT codes~\cite{emde2015iprt, emde2018iprt} with an aim of improving retrieval algorithms.
	Motivated by the CloudCT mission formulation---only the first of many to come in innovative passive cloud remote sensing---we develop herein a novel framework for 3D remote sensing of cloud properties using multi-view polarimetric measurements.
	
	\subsection{Why passive tomography?}
	From its etymology, the word ``tomography'' means a slice-by-slice recovery of an object's 3D internal structure using 2D projections of cumulative density. In the computer age, this task is termed {\em Computed Tomography} (CT) ~\cite{KakSlaney88}.
	Common medical CT approaches are transmission-based X-ray CT or 
	single-photon emission computed tomography (SPECT). There, 2D projections represent straight line-of-sight (LOS) integrals of the local X-ray opacity or nuclear marker density, respectively. In both imaging modalities, the inverse problem of recovering the medium content is {\em linear}~\cite{gordon1970algebraic}. 
	
	Biomedical imaging also involves CT modalities which are not based on linear projections. A prime example is Optical Diffusion Tomography (ODT)~\cite{arridge1999optical,boas2001imaging,arridge2009optical}, which uses non-ionizing near-infrared light. It is worth noting the work by Che et al.~\cite{che2018inverse} which departs from physics-based approaches into the realm of machine-learning.
	
	In ODT, a patient's organ is surrounded by a large number of point sources of pulsed isotropic near-infrared irradiation and a large number of time-resolving omnidirectional sensors. The organ transmits radiation diffusely, with very little absorption. Anomalous 3D absorbing or vacuous regions can be detected and assayed using {\em nonlinear} inverse diffusion spatiotemporal analysis, that relies on very high orders of scattering. The detected radiance is blurred, yielding limited 3D spatial resolution. However, ODT can yield sufficient diagnostic information, using non-ionising radiation. 
	
	Medical CT modalities generally use {\em active} radiation.  Active methods are also used for {\em local} atmospheric sensing or scatterers by radar and lidar. There, a  transmitter and receiver are generally collocated and signals are based on backscattering and time-resolved two-way transmission. Probing is solved per LOS using methods which are computationally relatively simple. However, the technology is expensive, horizontal sampling is generally very limited, and irradiance decays fast from the transmitter. Passive sensing is less expensive, uses minimal power, and can image wide swaths of Earth. Thus {\em global coverage mandates passive imaging from space}. Consequently, this paper focuses on derivation of 3D passive tomography of scatterer fields. 
	
	Passive remote sensing does not benefit from pulsed sources for echo-location. It should rely on multi-angular data. Linear CT models (analogous to medical Xray CT and SPECT) were used to study gas emission and absorption in 3D plumes in the vicinity of pollution sources~\cite{Hashmonay1999,Todd2001} or volcanoes~\cite{Kazahaya2008,Wright2008}. There, Rayleigh-scattered sunlight was transmitted through the gas to a spectrometer on a platform flying around the plume. Following the vision of Werner et al.~\cite{warner1986liquid}, Huang et al.~\cite{huang2008determination, huang2008cloud} used scanning microwave radiometers to reconstruct 2D slices of particle density in clouds based on its impact on local emissivity.
	
	Linear CT was also adapted by Garay et al.~\cite{garay2016tomographic} to characterize a smoke plume over water emanating from a coastal wild fire. There, the signal is sunlight scattered to space and detected by the Multi-angle Imaging Spectro-Radiometer (MISR) sensor~\cite{Diner1998} at nine viewing angles. The analysis in~\cite{garay2016tomographic} yields the direct transmission through the plume per LOS, from which linear CT analysis yields the plume density without using solar radiometers under the plume. 
	
	In general, however, retrieving atmospheric scatterer fields in 3D requires a full forward model of scattering in 3D. The model satisfies neither a direct transmission model of linear CT, nor the diffusion limit of ODT. In passive imaging of scatterers, the light source irradiating the atmosphere is the sun: uncontrolled, steady and mono-directional.  Aides et al.~\cite{AidesEtAl2013} formulated CT based on single-scattered light. Their forward model is based on sets of broken-ray paths, where light changes direction once from the sun to a sensor. 
	
	All the above atmospheric tomography methods assumed the medium to be optically thin enough for direct and once-scattered radiation to dominate the measured radiance. We depart radically from this assumption, drawing inspiration from the success of active ODT, though necessarily with a different forward model. 
	We formulate an inverse 3D RT problem for cloud tomography utilizing multi-view multi-spectral polarimetric images. In contrast to linear CT, the image formation model is nonlinear in the microphysical and density variables. Our approach seeks an optimal fit of droplet microphysical parameters. This is based on a computational 3D polarized RT forward model, the {\em vector Spherical Harmonics Discrete Ordinates Method} (vSHDOM)~\cite{evans1998spherical,doicu2013multi}.
	To this effect, we generalize our demonstrated iterative inversion approach~\cite{levis2015airborne,HolodovskyEtAl2016,levis2017multiple} to take advantage of polarimetric measurements.
	
	\subsection{Outline}
	In the next section, we cover basic cloud droplet optics using Mie scattering theory and the fundamentals of polarized 3D RT. The latter yields radiance which has a clear decomposition  into single- and multiply-scattered light. This decomposition supports the solution to the inverse problem at hand. We then lay out our 3D cloud tomography method where we target three basic microphysical properties, volumetrically.
	Necessary but tedious mathematical details are presented in the {\it Appendix}.
	Subsequently, the new 3D cloud tomographic capability is demonstrated on realistic synthetic clouds from a Large Eddy Simulation (LES) that provide ground truth for unambiguous retrieval error quantification.
	We conclude with a summary of our results and an outline of future developments, mostly looking toward CloudCT and other future space-based uses.
	
	\section{Background}
	\label{sec:background}
	
	This section describes bulk microphysical parameterization of scattering media, the polarimetric radiative transfer image formation (forward) model and the relation between them. The section also describes the coordinate systems in use (per-scatterer, imager and Earth frames). We further decompose the polarized radiance into single-scattered and high-order scattered components. These foundations are used in subsequent sections, to formulate tomographic recovery. 
	
	
	\subsection{Scatterer microphysical properties}
	\label{subsec:microphysics}
	
\begin{figure*}[t]
		\centering \includegraphics[width=0.9\textwidth]{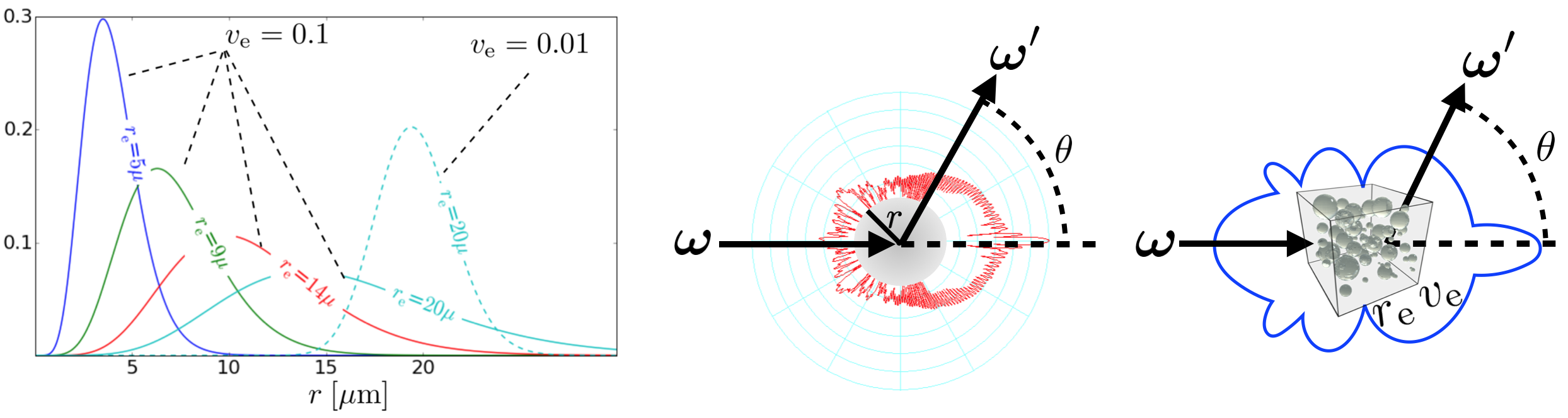}
		\caption{[Left] Normalized {\em Gamma}-distribution. The effective radius and variance dictate the centroid and width of the size-distribution. The limit of very low $v_{\rm e}$ approaches a mono-disperse distribution. [Center] Log-polar plot of the Mie phase-function $p_{11}$ induced by a single water sphere of radius $r$. [Right] Log-polar plot of the effective phase-function $\langle s_{\rm s} p_{11} \rangle_r / \sigma_{\rm s}$ induced by a small volume that includes particles of different sizes.}
		\label{fig:gamma}
	\end{figure*}

	In the lower atmosphere, cloud particles are droplets of liquid water that are very nearly spherical, having radius $r$. 
	They are however polydisperse, with a droplet size distribution denoted $n(r)$. 
	For most remote-sensing purposes, $n(r)$ is parameterized using an {\em effective radius} in $\mu$m and a dimensionless {\em variance}~\cite{hansen1971multiple}:
	\begin{eqnarray}
	r_{\rm e} =  \frac{ \int_0^{\infty} r^3 n(r) {\rm d}r}{\int_0^{\infty}  r^2 n(r) {\rm d}r}, \quad v_{\rm e} = \frac{\int_0^{\infty} \left(r {-} r_{\rm e}\right)^2 r^2n(r) {\rm d}r}{r_{\rm e}^2\int_0^{\infty} r^2 n(r) {\rm d}r}.
	\label{eq:reff_veff}
	\end{eqnarray}
	A commonly used parametric size distribution, having empirical support~\cite{hansen1971multiple} is the {\em Gamma}-distribution (Fig.~\ref{fig:gamma}):
	\begin{equation}
	n(r) = N \, c \, r^{(v_{\rm e}^{-1} \! {-}3)} \exp [- r / (r_{\rm e}v_{\rm e})],
	\label{eq:gamma}
	\end{equation}
	where we require $v_{\rm e} < 1/2$. Here $c = (r_{\rm e}v_{\rm e})^{(2{-}v_{\rm e}^{-1})} / \Gamma(v_{\rm e}^{-1} \! {-}2)$ is a normalization constant and
	\begin{equation}
	N = \int_0^{\infty} n(r) {\rm d}r
	\label{eq:N}
	\end{equation}
	is the droplet number concentration. 
	Let $\rho_{\rm w}$ be the density of liquid water. An important cloud characteristic is the water mass density or {\em Liquid Water Content} (LWC) per unit volume:
	\begin{equation}
	{\rm LWC} = \frac{4}{3} \pi \rho_{\rm w} \int_0^{\infty} r^3 n(r) {\rm d}r.
	\label{eq:LWC}
	\end{equation}
	It is expressed as LWC = $\nicefrac{4}{3} \, \pi \rho_{\rm w} r_{\rm e}^3 (1-v_{\rm e})(1-2v_{\rm e})$ for the Gamma distribution in \eqref{eq:gamma}.
	
	
	\subsection{Polarized light}
	\label{subsec:polarized}
	
	A light wave is associated with orthogonal components of a random electric wave, $E_1(t)$ and $E_2(t)$, where $t$ is time. 
	The components' direction unit vectors are respectively $\hat{\bf E}_1$ and $\hat{\bf E}_2$. 
	The wave propagates in direction ${\bm \omega} = \hat{\bf E}_1 \times \hat{\bf E}_2$. 
	It is convenient to define the polarized light state in terms of the Stokes~\cite{hansen1971multiple} vector ${\bf I} = \left(I, Q, U, V\right)^\top$. Each component of ${\bf I}$ expresses temporal expectation:
	\begin{eqnarray}
	I =  \langle E_{1}E^*_{1} + E_{2}E^*_{2}\rangle_t, \quad
	Q  =  \langle E_{1}E^*_{1} - E_{2}E^*_{2}\rangle_t,  \\
	U =  \langle E_{1}E^*_{2} + E_{2}E^*_{1}\rangle_t,  \nonumber \quad
	V  = {\rm i}\langle E_{1}E^*_{2} - E_{2}E^*_{1}\rangle_t, \nonumber
	\label{eq:stokes}
	\end{eqnarray}
	where ${\rm i} = \sqrt{-1}$. 
	Unpolarized intensity is $I$. 
	The degrees of polarization (DOP) and linear polarization (DoLP) are respectively defined as the ratios $\nicefrac{\sqrt{Q^2 + U^2 + V^2}}{I},~\nicefrac{\sqrt{Q^2 + U^2}}{I}$. The angle of linear polarization (AoLP) is $\nicefrac{1}{2}\,\tan^{-1}(U/Q)$.
	

	\subsection{Single scattering of polarized light}
	\label{subsec:single_scattering}
	
	Light interaction with a single particle is described by the total extinction cross-section $s_{\rm t}(r,\lambda)$, decomposed into scattering and absorption cross-sections, 
	respectively:
	\begin{eqnarray}
	s_{\rm t}(r,\lambda) = s_{\rm s}(r,\lambda) 
	+ s_{\rm a}(r,\lambda).
	\label{eq:single_crosssection}
	\end{eqnarray}
	In Mie scattering by spheres, introduced further on, we have
	\begin{eqnarray}
	\label{eq:Mie_crosssection}
	s_{\rm t}(r,\lambda) = \pi \, r^2 \, \mathcal{Q}_{\rm t}(2\pi \, r/\lambda), \quad
	s_{\rm s}(r,\lambda) = \pi \, r^2 \, \mathcal{Q}{\rm s}(2\pi \, r/\lambda), \quad s_{\rm a}(r,\lambda) = \pi \, r^2 \, \mathcal{Q}{\rm a}(2\pi \, r/\lambda) \nonumber
	\end{eqnarray}
	where $\mathcal{Q}{\rm t},\mathcal{Q}{\rm s}, \mathcal{Q}{\rm a}$ are dimensionless efficiency factors, which depend on the normalized size parameter $2\pi \, r/\lambda$. In the limit $r \gg \lambda$,
	$\mathcal{Q}{\rm t} \approx 2$. Furthermore, when  $s_{\rm s}(r,\lambda) \gg s_{\rm a}(r,\lambda)$, then $\mathcal{Q}{\rm s} \approx 2$ and $\mathcal{Q}{\rm a} \approx 0$.
	
	Define size-weighted average over some function $a(r)$ by\footnote{
		The size integral of \eqref{eq:r_average} is in practice terminated at $r_{\rm max}$ = 70~$\mu$m.
	}
	\begin{eqnarray}
	\langle  a \rangle_r = \frac{1}{N}\int_0^{\infty} a(r) n(r) {\rm d} r.
	\label{eq:r_average}
	\end{eqnarray}
	Note that we use here an approximation, commonly used in multi-spectral remote sensing, of a single rendering with spectrally-averaged optical properties.  
	The material optical properties can furthermore be approximated, in the absence of molecular absorption, by using a single wavelength for each spectral band.  
	This is valid if wavelength dependencies within a spectral band are weak, a condition met when narrow bands are considered. 
	Macroscopic optical cross-sections are then expressed as weighted averages\footnote
	{Aggregating scattered properties in \eqref{eq:bulk_cs} rather than electric fields holds for scatterer populations that are in each other's far field (i.e., are $\gg \lambda$ apart)~\cite{hansen1971multiple}.}
	\begin{eqnarray}
	\sigma_{\rm t}(\lambda) {=} \langle s_{\rm t}(r,\lambda) \rangle_r,~
	\sigma_{\rm s}(\lambda) {=}  \langle  s_{\rm s}(r,\lambda) \rangle_r,~
	\sigma_{\rm a}(\lambda)  {=}  \langle  s_{\rm a}(r,\lambda) \rangle_r.
	\label{eq:bulk_cs}
	\end{eqnarray}
	Throughout the text, dependency on $\lambda$ is generally omitted for simplicity; however, it is used at specific points as needed. 
	
	Scattering, as a fraction of the overall interaction~\cite{marshak20053d}, is expressed by the dimensionless {\em single scattering albedo}
	\begin{eqnarray}
	\varpi = \frac{\sigma_{\rm s}}{\sigma_{\rm t}}.
	\label{eq:ssa}
	\end{eqnarray}
	The extinction coefficient (or optical density) is denoted by $\beta$. 
	Following Eqs.~[\ref{eq:N},\ref{eq:LWC},\ref{eq:bulk_cs}], $\beta = N \sigma_{\rm t}$ is expressed in terms of the LWC as~\cite{chylek1978extinction}
	\begin{eqnarray}
	\beta = \frac{{\rm LWC}}{\frac{4}{3}\pi \rho_w \langle r^3 \rangle_r} \sigma_{\rm t} = {\rm LWC} \cdot \tilde{\sigma}_{\rm t}.
	\label{eq:extinct}
	\end{eqnarray}
	Here, $\tilde{\sigma}_{\rm t}$ is the mass extinction coefficient (in units of $\nicefrac{{\rm m}^2}{{\rm g}}$).
	
	Let ${\bm \omega}$ and ${\bm \omega}'$ be the unitary incident and scattered ray direction vectors respectively in Fig.~\ref{fig:gamma}. 
	Single-scattering geometry is defined by the local coordinate system of the incoming beam's electric fields. 
	As stated above, the electric field of incoming light is decomposed into components along orthogonal directions. 
	We set them as
	\begin{equation}
	{\bf E}_1 \propto {\bm \omega} \times {\bm \omega}', \quad 
	{\bf E}_2 \propto {\bf E}_1 \times {\bm \omega}.
	\end{equation}
	The scattering angle is $\theta = \cos^{-1}({\bm \omega} {\cdot} {\bm \omega}')$. 
	The angular redistribution of singly-scattering light from a sphere of is defined by the $4{\times}4$ dimensionless {\em Mueller} matrix ${\bf P}_{\rm s}(\theta, r)$. 
	The macroscopic {\em phase matrix} is the size-weighted average
	\begin{eqnarray}
	{\bf P}(\theta) = \frac{\langle  s_{\rm s}(r) {\bf P}_{\rm s}(\theta, r) \rangle_r}{\sigma_{\rm s}}.
	\label{eq:bulk_p}
	\end{eqnarray}
	For spherical (or just randomly-oriented) particles, the phase-matrix ${\bf P}(\theta)$ takes the following symmetric form~\cite{hansen1971multiple}
	\begin{eqnarray}
	{\bf P} \left(\theta \right) = 
	\begin{bmatrix} 
	p_{11} \left(\theta \right)  & p_{21} \left(\theta \right)  & 0 & 0   \\
	p_{21} \left(\theta \right)  & p_{22} \left(\theta \right)  & 0 & 0   \\
	0 & 0  & p_{33} \left(\theta \right)  & -p_{43} \left(\theta \right)  \\
	0 & 0  & p_{43} \left(\theta \right)  &  p_{44} \left(\theta \right)  \\
	\end{bmatrix},
	\label{eq:phasemat}
	\end{eqnarray}
	where $p_{11}$ is the (unpolarized) scattering phase-function. 
	In single-scattering of unpolarized incident sunlight, the DoLP of scattered light amounts to the ratio $|p_{21}| / p_{11}$.
	
	
	\subsubsection{Rayleigh scattering}
	The Rayleigh model describes light scattering by particles much smaller than the wavelength. The Rayleigh phase matrix takes the following form~\cite{chandrasekhar1950radiative}
	\begin{eqnarray}
	{\bf P}_{\rm Rayl} \left(\theta \right) = 
	\begin{bmatrix} 
	\frac{3}{4}\left(1+\cos^2\theta\right)  & -\frac{3}{4}\sin^2\theta  & 0 & 0   \\
	-\frac{3}{4}\sin^2\theta  & \frac{3}{4}\left(1+\cos^2\theta\right)  & 0 & 0   \\
	0 & 0  & \frac{3}{2}\cos\theta  & 0 \\
	0 & 0  & 0 &  \frac{3}{2}\cos\theta  \\
	\end{bmatrix}.
	\label{eq:pol_rayl}
	\end{eqnarray}
	The single-scattering DoLP due to air molecules is then
	\begin{eqnarray}
	{\rm DoLP}_{\rm Rayl}(\theta) = \frac{\sin^2\theta}{1+ \cos^2\theta}.
	\label{eq:dolp_rayl}
	\end{eqnarray}
	According to \eqref{eq:dolp_rayl} a maximum DoLP is attained at single-scattering angle $\theta = 90^{\circ}$.
	
	
	\subsubsection{Mie scattering}
	Mie theory describes how light interacts with a spherical particle of size comparable to $\lambda$~\cite{bohren2008absorption}. 
	Denote $\mu{=}\cos\theta$. Mie scattering is defined in terms of complex-valued amplitude scattering functions\footnote{
		For a full mathematical description, see \cite{bohren2008absorption}.
	} 
	$S_1(\mu), S_2(\mu)$, which correspond to scattering of the $E_1, E_2$ electric field components. 
	Scattering of the Stokes vector ${\bf I}$ is described by the phase matrix ${\bf P}_{\rm Mie}(\mu)$, which is fully defined by six matrix components:
	\begin{eqnarray}
	p^{\rm Mie}_{11} = \frac{\varrho}{2}\left(S_1 S_1^* + S_2 S_2^*\right), \nonumber \quad
	p^{\rm Mie}_{12} = \frac{\varrho}{2}\left(S_1 S_1^* - S_2 S_2^*\right), \nonumber \\
	p^{\rm Mie}_{22} = \frac{\varrho}{2}\left(S_1 S_1^* + S_2 S_2^*\right), \nonumber \quad
	p^{\rm Mie}_{33} = \frac{\varrho}{2}\left(S_1 S_2^* + S_2 S_1^*\right), \nonumber \\
	p^{\rm Mie}_{43} = \frac{\varrho}{2}\left(S_1 S_2^* - S_2 S_1^*\right), \nonumber \quad
	p^{\rm Mie}_{44} = \frac{\varrho}{2}\left(S_1 S_2^* + S_2 S_1^*\right).
	\label{eq:pol_mie}
	\end{eqnarray}
	Here, $\varrho$ is a normalization constant, set to satisfy $\frac{1}{2}\int_{-1}^1  p^{\rm Mie}_{11} (\mu ) {\rm d} \mu = 1$.
	
\begin{figure*}[t]
		\centering \includegraphics[width=0.9\textwidth]{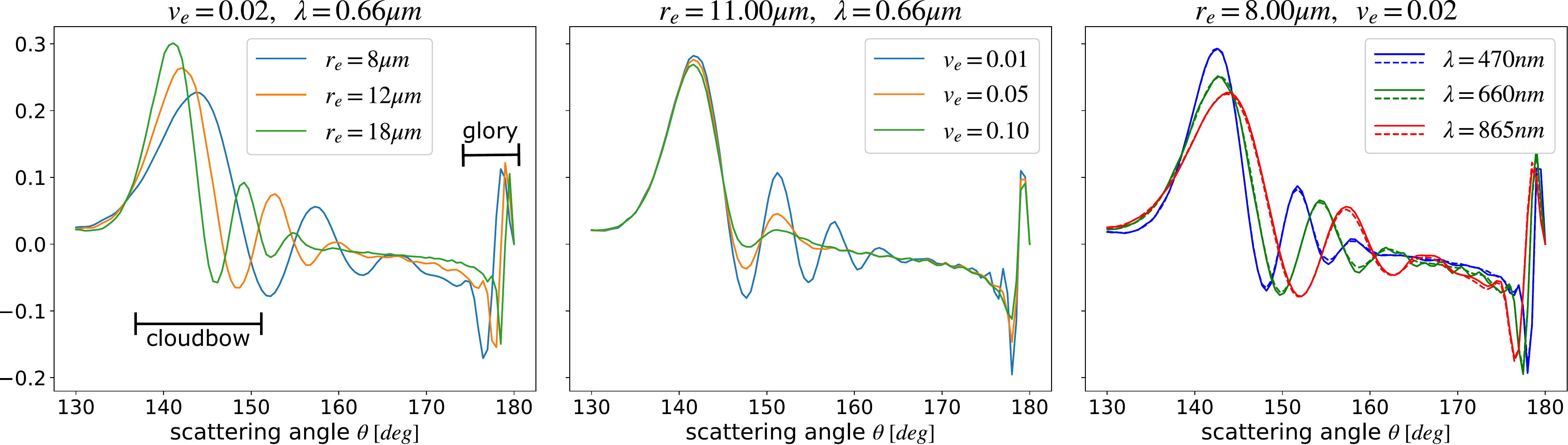}
		\caption{Normalized phase matrix element $-p^{\rm Mie}_{12}/p^{\rm Mie}_{11}$ around the cloud-bow and glory regions. For highly disperse droplet distributions (large $v_e$) the secondary lobes of the cloud-bow ($\theta \sim 140^{\circ}$) and glory ($\theta \sim 180^{\circ}$) diminish. The main cloud-bow peak is slightly sensitive to $\lambda$ or $v_{\rm e}$. The side-lobe angles are more sensitive to $\lambda$ and $r_{\rm e}$. The side-lobe amplitude is sensitive to $v_{\rm e}$. This cloud-bow signal is helpful for retrievals of $r_{\rm e}$. [Right plot] Solid lines indicate monochromatic light. Dashed lines indicate spectral averaging over a 100~nm bandwidth, which is more than double any of the spectral bands considered further on.}
		\label{fig:mie_cloudbow_glory}
	\end{figure*}

	Mie scattering due to water droplets is peaked at specific angles. 
	For a single droplet or monodisperse material, ${\bf P}^{\rm Mie}$ has sharp scattering lobes at angles that depend on the droplet's $\nicefrac{r}{\lambda}$ ratio. 
	A macroscopic voxel contains droplets in a range of radii $r$, smoothing the scattering lobes. 
	The smoothing effect depends on $v_{\rm e}$ (Fig.~\ref{fig:mie_cloudbow_glory}) and, to a far lesser extent, the spectral bandwidth
	(Fig.~\ref{fig:mie_cloudbow_glory}). 
	Two angular domains that stand out for remote-sensing purposes are the cloud-bow ($\theta \in [135^{\circ}, 155^{\circ}]$) and glory ($\theta \in [175^{\circ}, 180^{\circ}]$). 
	Both domains have peaks that are sensitive to the droplet microphysical parameters, and are significantly polarized (i.e., peaks are visible in the $p_{12}^{\rm Mie}$ component). 
	The latter fact renders these peaks distinguishable in the presence of a multiply-scattered signal component.
	
	
	\subsection{Multiple scattering of polarized light} 
	\label{subsec:multiple_scattering}
	
\begin{figure*}[t]
		\centering \includegraphics[width=0.75\textwidth]{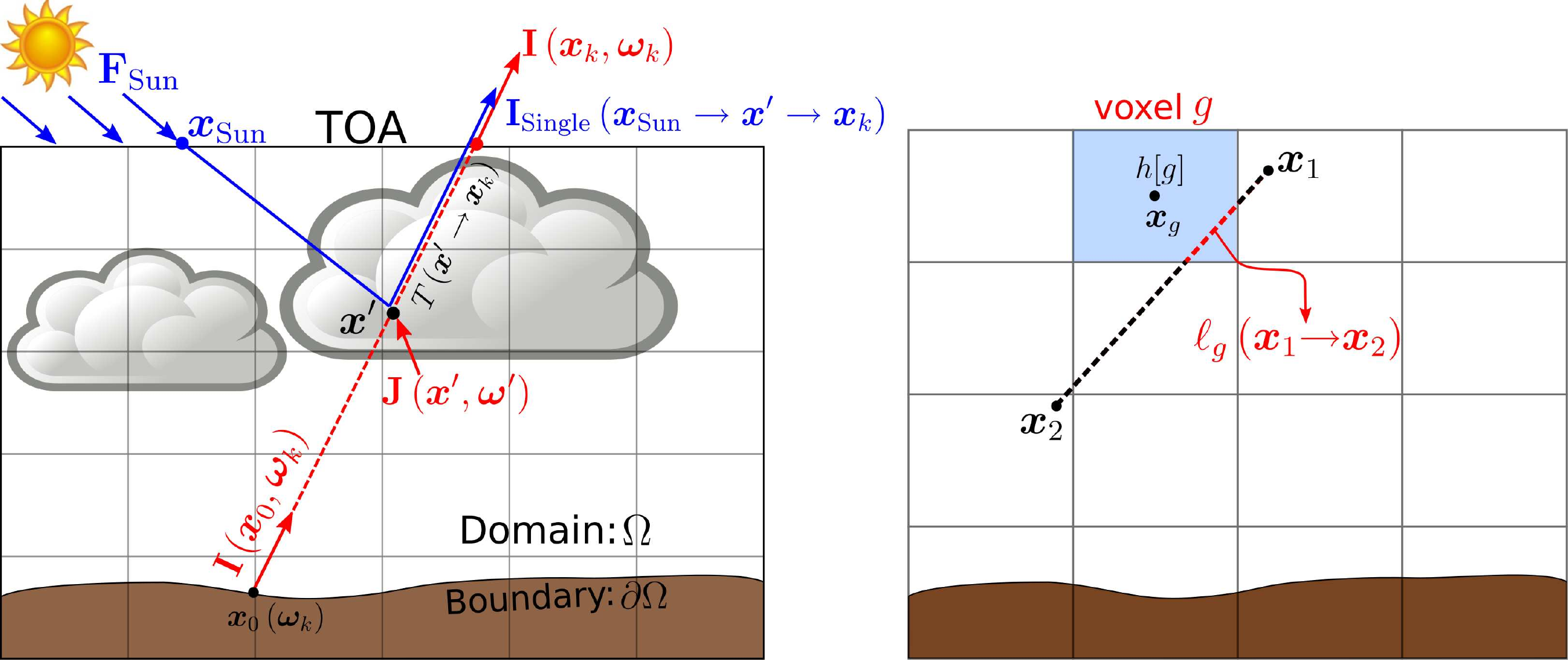}
		\caption{[Left] Light scatters in the medium, generally multiple times, creating a partially polarized (vector) scatter field ${\bf J}$ (\eqref{eq:J}). Integration yields the partially polarized (vector) light field ${\bf I}$ (\eqref{eq:rte_integral}). Here $I({\bm x}_k, {\bm \omega}_k)$ is a pixel measurement at the TOA and ${\bf I}_{\rm Single}$ is the single-scattered contribution from ${\bm x}'$ [Right] Ray tracing of a line-integral over a discretized voxel field $h[g]$ (zero-order interpolation).}
		\label{fig:domain}
	\end{figure*}

	The {\em Radiative Transfer Equation} (RTE)~\cite{chandrasekhar1950radiative} describes multiple scattering interactions of monochromatic partially polarized light within a medium. 
	Transmittance between two points ${\bm x}_1, {\bm x}_2$ is
	\begin{eqnarray}
	\label{eq:T}
	T \left({\bm x}_1 {\rightarrow} {\bm x}_2\right) =
	\exp{\left[-\int_{{\bm x}_1}^{{\bm x}_2}
		\! \beta({\bm x}){\rm d}{\bm x}\right]}.
	\end{eqnarray}
	An atmospheric domain $\Omega$ has boundary $\partial \Omega$. 
	The intersection of $\partial \Omega$ with a ray originating at point ${\bm x}$ in direction $-{\bm \omega}$ (Fig.~\ref{fig:domain}) is denoted ${\bm x}_0 ({\bm x},\omega)$. 
	Denote the Stokes vector field as ${\bf I} \xw$. 
	Then ${\bf I}({\bm x}_0, {\bm \omega})$ is the Stokes vector of radiation which propagates 
	in direction ${\bm \omega}$ at boundary point ${\bm x}_0({{\bm x},\bm \omega})$. 
	The non-emissive forward RT model~\cite{chandrasekhar1950radiative} couples ${\bf I}\xw$ to a vector {\em source field} ${\bf J}\xw$ (Fig.~\ref{fig:domain}) by
	\begin{eqnarray}
	&& {\bf I} \xw  =  {\bf I}({\bm x}_0, {\bm \omega}) T\left({\bm x}_0 {\rightarrow} {\bm x}\right)
	+ 
	\int_{{\bm x}_0}^{\bm x}  {\bf J}({\bm x}',{\bm \omega}) \beta({\bm x}') T\left({\bm x}'  {\rightarrow} {\bm x}\right) {\rm d}{\bm x}',  \label{eq:rte_integral}  \\
	&&{\bf J} \xw  =  \frac{\varpi({\bm x})}{4\pi} \int_{4\pi}
	{\bf P}\left({\bm x},{\bm \omega}{\cdot}{\bm \omega'}\right) {\bf I} \left({\bm x},{\bm \omega'}\right)   {\rm d}{\bm \omega}'.
	\label{eq:J}
	\end{eqnarray}
	Equations~[\ref{eq:rte_integral}-\ref{eq:J}] are solved numerically, either directly with an explicit solver~\cite{doicu2013multi} or indirectly using a Monte-Carlo path tracer~\cite{mayer2009radiative}. 
	We use vSHDOM~\cite{doicu2013multi} to simulate scattered Stokes components of a realistic atmosphere, having both Mie and Rayleigh scattering due to water droplets and air molecules. 
	
	Multiple scattering interactions are defined using two coordinate systems. 
	Local {\em scatterer coordinates} are set by \mbox{$(\hat{\bf E}_1, \hat{\bf E}_2)$}. 
	Stokes measurements in satellites, however, are defined in {\em Meridional} coordinates. 
	Let $\hat{{\bf z}}$ denote the zenith direction vector at every point on Earth. 
	In meridian coordinates, the electric field components are defined by direction vectors
	\begin{eqnarray}
	\hat{{\bf m}}_1  =  \frac{\hat{{\bf z}} \times {\bm \omega}}{ \| \hat{{\bf z}} \times {\bm \omega}\|}, \quad
	\hat{{\bf m}}_2 = {\bm \omega} \times \hat{{\bf m}}_1.
	\end{eqnarray}
	Each pixel-scale Stokes measurement is described by a coordinate system defined by $\hat{{\bf m}}_1$ and $\hat{{\bf m}}_2$. 
	The transformation between the two coordinate systems amounts to a multiplication of ${\bf I}$ by a Mueller rotation matrix.
	
	Sampling ${\bf I}\xw$ at the location of each camera and direction of each camera pixel yields the measured Stokes vector. 
	A measurement $k$ is done at the camera position ${\bm x}_k$, LOS direction ${\bm \omega}_k$, and wavelength $\lambda_k$ (Fig.~\ref{fig:domain}). 
	Thus,  Eqs.~[\ref{eq:rte_integral}-\ref{eq:J}] yield the pixel measurement model
	\begin{eqnarray}
	{\bf I}[k] = {\bf I} \left({\bm x}_0, {\bm \omega}_k \right) T\left({\bm x}_0 {\rightarrow} {\bm x}_k\right) 
	+ 
	\int_{{\bm x}_0}^{{\bm x}_k} {\bf J}\left({\bm x}',{\bm \omega}_k\right) \beta({\bm x}') T\left({\bm x}'  {\rightarrow} {\bm x}_k\right) {\rm d}{\bm x}'. 
	\label{eq:forward_stokes}
	\end{eqnarray}
	
	
	\subsection{Single-scattering separation}
	\label{subsec:singscat_sep}
	It is often convenient to separate the single-scattering contribution from the rest of the radiance field~\cite{nakajima1988algorithms}. 
	The solar irradiance at the top of the atmosphere (TOA) is $F_{\rm Sun}$. 
	It is unoplarized, thus corresponds to a Stokes vector ${\bf F}_{\rm Sun}{=}\left( F_{\rm Sun},~0,~0,~0 \right)^\top$. 
	The Sun is modeled as an ideal directional source with direction ${\bm \omega}_{\rm Sun}$. 
	A solar ray heading to point ${\bm x}$ intersects the TOA at point ${\bm x}_{\rm Sun}$. 
	The solar transmittance is given by $T\left({\bm x}_{\rm Sun} {\rightarrow} {\bm x}\right)$. 
	Let $\delta$ denote {\em Dirac's delta}. 
	Thus, ${\bf I}$ can be written as a sum of the {\em diffuse} component ${\bf I}_{\rm d}$, and direct solar component:
	\begin{equation}
	{\bf I} \xw = {\bf I}_{\rm d} \xw  + \delta \left({\bm \omega} {-} {\bm \omega}_{\rm Sun} \right){\bf F}_{\rm Sun}  T\left({\bm x}_{\rm Sun} {\rightarrow} {\bm x}\right) .
	\label{eq:I_seperation}
	\end{equation}
	Inserting [\ref{eq:I_seperation}] into [\ref{eq:J}] yields 
	\begin{equation}
	\label{eq:J_seperation}
	{\bf J} \xw = {\bf J}_{\rm d} \xw  + 
	\frac{\varpi({\bm x})}{4\pi} 
	{\bf P}\left({\bm x},{\bm \omega}{\cdot}{\bm \omega}_{\rm Sun}\right){\bf F}_{\rm Sun}   T\left({\bm x}_{\rm Sun} {\rightarrow} {\bm x}\right), 
	\end{equation}
	where 
	\begin{equation}
	{\bf J}_{\rm d} \xw = \frac{\varpi({\bm x})}{4\pi} \int_{4\pi}
	{\bf P}\left({\bm x},{\bm \omega}{\cdot}{\bm \omega'}\right) {\bf I}_{\rm d} \left({\bm x},{\bm \omega'}\right) {\rm d}{\bm \omega}' .
	\label{eq:J_d}
	\end{equation} 
	Consider Fig.~\ref{fig:domain}. Denote a {\em broken-ray} path of direct sunlight which undergoes single scattering at ${\bm x}'$, then reaches the camera:
	\begin{equation}
	{\bm x}_{\rm Sun} {\rightarrow} {\bm x}' {\rightarrow} {\bm x}_k.
	\label{eq:broken_path}
	\end{equation}
	It projects in direction ${\bm \omega}_k$ to pixel at ${\bm x}_k$, thus contributing to the  measurement ${\bf I}({\bm x}_k, {\bm \omega}_k)$. 
	Using Eqs.~[\ref{eq:rte_integral},\ref{eq:J_seperation}], the single-scattered contribution from ${\bm x}'$ is 
	\begin{eqnarray}
	{\bf I}_{\rm Single}\left({\bm x}_{\rm Sun} {\rightarrow} {\bm x}' {\rightarrow} {\bm x}_k \right) = 
	\frac{\varpi({\bm x}')}{4\pi} \beta ({\bm x}') {\bf P}\left({\bm x}',{\bm \omega}_k{\cdot}{\bm \omega}_{\rm Sun}\right){\bf F}_{\rm Sun}  T\left({\bm x}_{\rm Sun} {\rightarrow} {\bm x}'\right)  T\left({\bm x}' {\rightarrow} {\bm x}_k\right). 
	\label{eq:single_scatter_x} 
	\end{eqnarray}
	Thus, the entire single-scattered signal accumulates contributions along the LOS
	\begin{equation}
	{\bf I}_{\rm Single}\left({\bm x}_k \right) = \int_{{\bm x}_0}^{{\bm x}_k} {\bf I}_{\rm Single}\left({\bm x}_{\rm Sun} {\rightarrow} {\bm x}' {\rightarrow} {\bm x}_k \right) {\rm d}{\bm x}'.
	\label{eq:single_scatter}
	\end{equation}
	
	\subsection{Ray tracing}
	\label{subsec:raytrace}
	Ray tracing computes a function over a straight line through a 3D domain. 
	A common operation is path-integration (e.g. Eqs.~[\ref{eq:T},\ref{eq:rte_integral}]). 
	Let $h({\bm x})$ be a continuous field. Define a grid of discrete points \mbox{${\bm x}_g,~\text{where}~g=1,2,...,N_{\rm grid}$}. 
	Denote $h[g] = h({\bm x}_g)$. 
	A path-integral over $h({\bm x})$ is numerically computed using an interpolation kernel $K$
	\begin{eqnarray}
	\label{eq:interpolation_kernel}
	\int_{{\bm x}_1}^{{\bm x}_2}
	\! h({\bm x}){\rm d}{\bm x} = \sum \limits_{g=1}^{N_{\rm grid}} h[g] \int_{{\bm x}_1}^{{\bm x}_2} K\left({\bm x} - {\bm x}_g\right){\rm d}{\bm x}.
	\end{eqnarray}
	For zero-order interpolation (i.e., voxel grid), \eqref{eq:interpolation_kernel} degenerates to 
	\begin{eqnarray}
	\label{eq:numerical_integration}
	\int_{{\bm x}_1}^{{\bm x}_2}
	\! h({\bm x}){\rm d}{\bm x} = \sum \limits_{g=1}^{N_{\rm grid}} h[g] \ell_g \left({\bm x}_1 {\rightarrow} {\bm x}_2 \right), 
	\end{eqnarray}
	where $\ell_g \left({\bm x}_1 {\rightarrow} {\bm x}_2 \right)$ is the intersection of the path with voxel $g$ (Fig.~\ref{fig:domain}). 
	For voxel indices $g$ that do not intersect the path ${\bm x}_1 {\rightarrow} {\bm x}_2$, the value of $\ell_g \left({\bm x}_1 {\rightarrow} {\bm x}_2 \right)$ is 0. 
	

	section{Cloud Tomography}
	\label{sec:tomography}
	So far, we described the forward (image-formation) model, i.e., how images are formed, given cloud properties. 
	In this work, we formulate a novel inverse tomographic problem of recovering the unknown cloud microphysical properties, volumetrically. 
	In voxel $g$, the vector of unknown parameters is $\left({\rm LWC}[g],~r_{\rm e}[g],~v_{\rm e}[g] \right)$. 
	The unknown microphysical parameters are concatenated to a vector of length $3N_{\rm grid}$
	\begin{equation}
	{\bf \Theta} = \big(...,{\rm LWC}[g],~r_{\rm e}[g],~v_{\rm e}[g],...\big)^\top, \quad 1\leq g \leq N_{\rm grid}.
	\label{eq:param_vec}
	\end{equation}
	Neglecting circular polarization, each pixel measures a Stokes vector, ${\bf y}_{\bf I}=\big(y_{\rm I}, y_{\rm Q}, y_{\rm U}\big)$ at $N_{\lambda}$ wavelengths. 
	Let $N_{\rm views}$ and $N_{\rm pix}$ denote the number of view points and camera pixels. The total number of Stokes measurements is thus $N_{\rm meas} {=} N_{\lambda} N_{\rm views} N_{\rm pix}$. 
	The measurement vector of length $3N_{\rm meas}$ is expressed as
	\begin{equation}
	{\bf y} = \big( {\bf y}_{\bf I}[1],....,{\bf y}_{\bf I}[N_{\rm meas}]\big)^{\top}.
	\label{eq:meas_vec}
	\end{equation}
	In this section, we formulate the use of measurements ${\bf y}$ (multi-view, multi-pixel, multi-spectral, polarimetric measurements) for tomographic retrieval of ${ \bf \Theta}$ (3D volumetric cloud density and microphysics). 
	It is worth mentioning at this point that Stokes components are not measured directly. 
	Rather, they are computationally retrieved from measurements of different polarization states (see {\it Appendix} for the AirMSPI measurement model). 
	
	
	\subsection{Polarimetric information}
	\label{subsec:information}
\begin{figure}[t]
		\centering \includegraphics[width=0.3\linewidth]{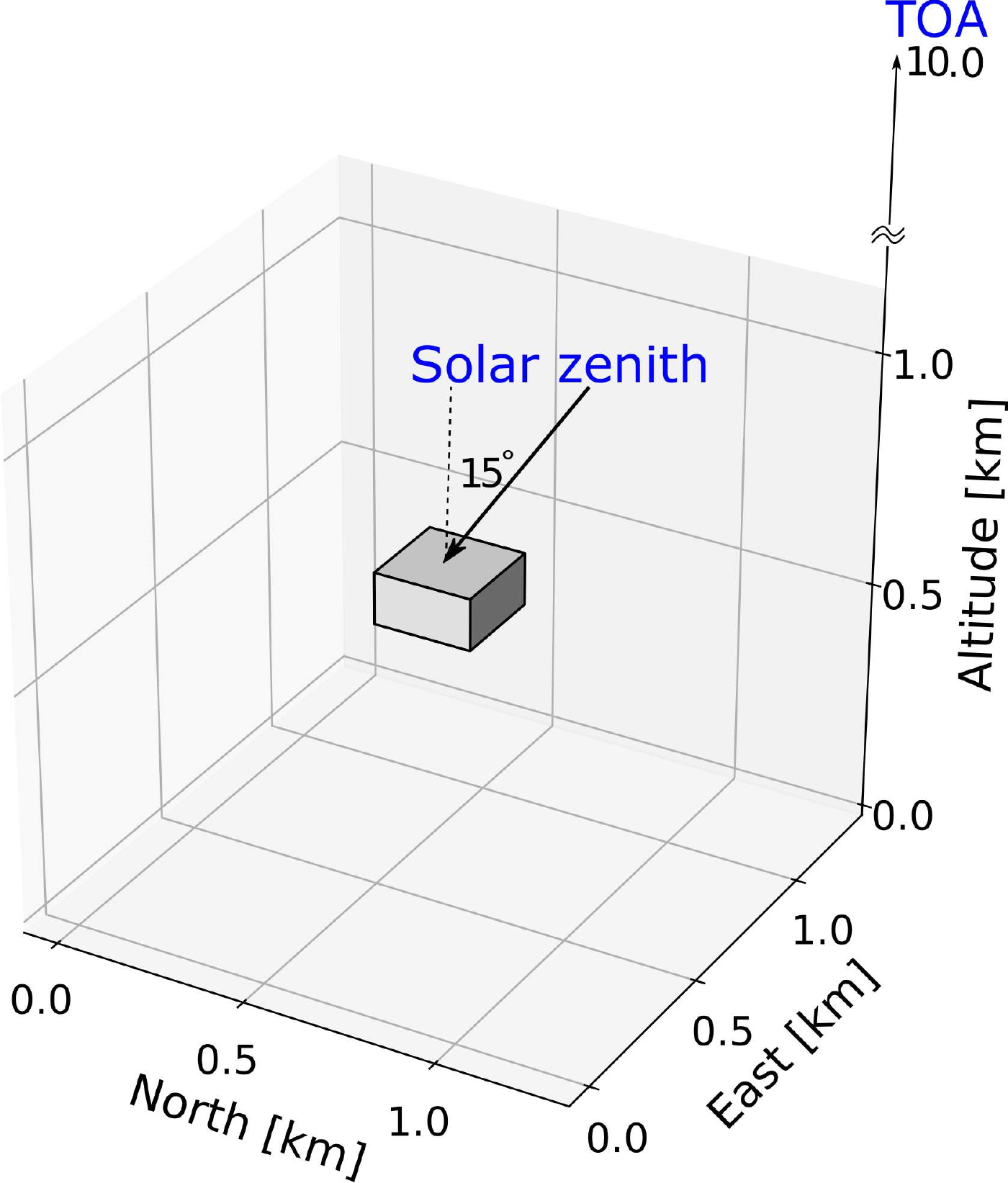}
		\caption{A homogeneous cubic cloud illuminated with solar radiation at a zenith angle of $15^\circ$ off-nadir. The solar azimuth angles are $\phi_0 = [0.0^\circ, 67.5^\circ]$. The outgoing Stokes vector ${\bf I}$ is simulated at AirMSPI resolution and wavelengths, with AirMSPI measuring along a North-bound track.} 
		\label{fig:cuboid}
	\end{figure}

\begin{figure}[t]
		\centering \includegraphics[width=0.7\linewidth]{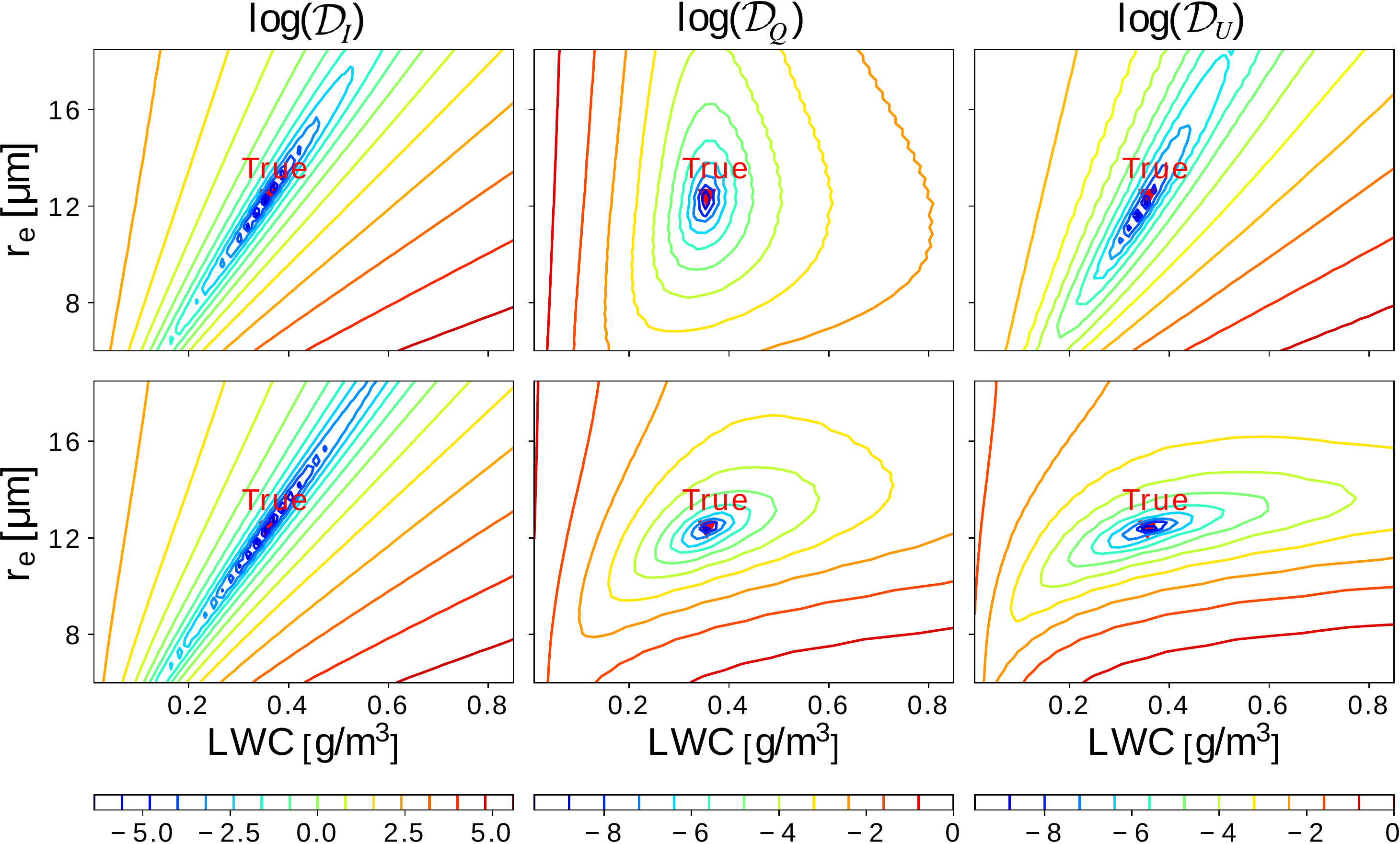}
		\caption{Logarithm of the 2D cost manifolds for a 2-parameter homogeneous cubic cloud (Fig.~\ref{fig:cuboid}). Each column of plots corresponds to the cost of the different Stokes components in Eqs.~[\ref{eq:minimization21}-\ref{eq:minimization22}]. Each row of plots corresponds to a different Solar azimuth angle $\phi_0$.} 
		\label{fig:loss_manifold}
	\end{figure}

	To make an initial assessment of the sensitivity of polarimetric measurements, we simulate a simple homogeneous cubic cloud (Fig.~\ref{fig:cuboid}), parameterized by two microphysical parameters: $({\rm LWC}, r_{\rm e})$. 
	Back-scattered Stokes measurements are taken at the TOA for angles and wavelengths sampled by the Airborne Multi-angle Spectro-Polarimetric Imager (AirMSPI)~\cite{diner2013airborne}.
	Define $I[k], U[k], Q[k]$ as simulated Stokes components at 
	measurement index $k$. 
	Define a cost function for each of the Stokes components
	\begin{eqnarray}
	\mathcal{D}_{\rm I}\left({\rm LWC}, r_{\rm e}\right) &=&  \sum_{k=1}^{N_{\rm meas}}  \left(I [k] - y_{\rm I}[k] \right)^2, \label{eq:minimization21} \\
	\mathcal{D}_{\rm Q}\left({\rm LWC}, r_{\rm e}\right) &=& \sum_{k=1}^{N_{\rm meas}}  \left(Q [k] - y_{\rm Q}[k] \right)^2, \\
	\mathcal{D}_{\rm U}\left({\rm LWC}, r_{\rm e}\right) &=&  \sum_{k=1}^{N_{\rm meas}}  \left(U [k] - y_{\rm U}[k] \right)^2,
	\label{eq:minimization22}
	\end{eqnarray}
	where we hold $v_{\rm e}$ constant.  Equations~[\ref{eq:minimization21}-\ref{eq:minimization22}] are 2D manifolds. 
	Figure~\ref{fig:loss_manifold} plots the cost manifolds for different solar azimuth angles, $\phi_0$. 
	While there is an ambiguity between ${\rm LWC}$ and $r_{\rm e}$ when relying on $\mathcal{D}_{\rm I}$, there are better defined minima for $\mathcal{D}_{\rm Q}$ and $\mathcal{D}_{\rm U}$. This indicates that polarization measurements carry valuable information. 
	
	
	\subsection{Inverse problem formulation}
	\label{subsec:inv_formulation}
	Denote ${\bf I}_{\bf \Theta}$ as the image formation model. 
	Tomography can be formulated as minimization of a data-fit function. 
	We preform 
	\begin{eqnarray}
	\hat{{\bf \Theta}} = \underset{{\bf \Theta}}{\arg\min}~ \mathcal{D} \left({\bf I}_{\bf \Theta}, {\bf y} \right) = \underset{{\bf \Theta}}{\arg\min} \left({\bf I}_{\bf \Theta} - {\bf y} \right)^{\top} {\bf \Sigma}^{-1} \left({\bf I}_{\bf \Theta} - {\bf y} \right), 
	\label{eq:minimization0}
	\end{eqnarray}
	Here ${\bf \Sigma}$ is related to the co-variance of the measurement noise. 
	For brevity, we omit the subscript ${\bf \Theta}$ but remember that 
	\begin{equation}
	{\bf I} {\equiv} {\bf I }_{\bf \Theta}, ~~ {\bf J} {\equiv} {\bf J}_{\bf \Theta}, ~~ \beta {\equiv} \beta_{\bf \Theta}, ~~ \varpi {\equiv} \varpi_{\bf \Theta}, ~~ {\bf P} {\equiv} {\bf P}_{\bf \Theta}, ~~ T {\equiv} T_{\bf \Theta}.
	\label{eq:shorthand}
	\end{equation}
	Assuming noise in different pixels, wavelengths and angles is uncorrelated, \eqref{eq:minimization0} degenerates to  
	\begin{equation}
	\hat{{\bf \Theta}} = \underset{{\bf \Theta}}{\arg\min} \sum \limits_{k=1}^{N_{\rm meas}} \left({\bf I}[k] - {\bf y}_{\bf I}[k] \right)^{\top} {\bf R}^{-1} \left({\bf I}[k] - {\bf y}_{\bf I}[k] \right).
	\label{eq:minimization}
	\end{equation}
	The matrix ${\bf R}$ depends on the particular sensor technology. 
	Description of ${\bf R}$, tailored to the AirMSPI sensor, is detailed in the {\it Appendix}. 
	
	We solve \eqref{eq:minimization} by a gradient-based approach. 
	The gradient with respect to the unknown parameters ${\bf \Theta}$ is 
	\begin{eqnarray}
	{\bf \nabla}_{\bf \Theta} \mathcal{D} \left({\bf I}_{\bf \Theta}, {\bf y}\right) = 2\sum \limits_{k=1}^{N_{\rm meas}} \big( {\bf I}[k]-{\bf y}_{\bf I}[k] \big)^{\top} {\bf R}^{-1} {\bf \nabla}_{\bf \Theta} {\bf I}[k].
	\label{eq:grad}
	\end{eqnarray}
	The term ${\bf \nabla}_{\bf \Theta} {\bf I}[k]$ is the {\em Jacobian} of the sensing model. 
	Equation [\ref{eq:grad}] is used to formulate an update rule for an iterative optimization algorithm 
	\begin{equation}
	{\bf \Theta}_{b+1} = {\bf \Theta}_{b} - \chi_b \nabla_{\bf \Theta} \mathcal{D}\left({\bf I}_{\bf \Theta},{\bf y} \right),
	\label{eq:update_rule}
	\end{equation}
	where $b$ denotes the iteration index and $\chi_b$ is a scalar. 
	We use L-BFGS~\cite{zhu1997algorithm} for numerical optimization that, in particular, determines adaptively the value of $\chi_b$. 
	One approach to computing the gradient $\nabla_{\bf \Theta} \mathcal{D}$ is the {\em Adjoint RTE}~\cite{doicu2019linearizations,martin2014adjoint}. 
	Due to the recursive nature of the RTE, computing the gradient through the exact Jacobian ${\bf \nabla}_{\bf \Theta} {\bf I}[k]$ is computationally expensive. 
	In the following sections, we derive a method to make the computation of the gradient tractable and efficient. 
	We do that by approximating the Jacobian $\nabla_{\bf \Theta} {\bf I}$ in a tractable way, using a two-step iterative algorithm~\cite{levis2015airborne,levis2017multiple}.
	
	
	\subsection{Iterative solution approach}
	\label{subsec:iterative_sol}
\begin{figure}[t]
		\def\svgwidth{\columnwidth}
		\centering \includegraphics[width=0.6\linewidth]{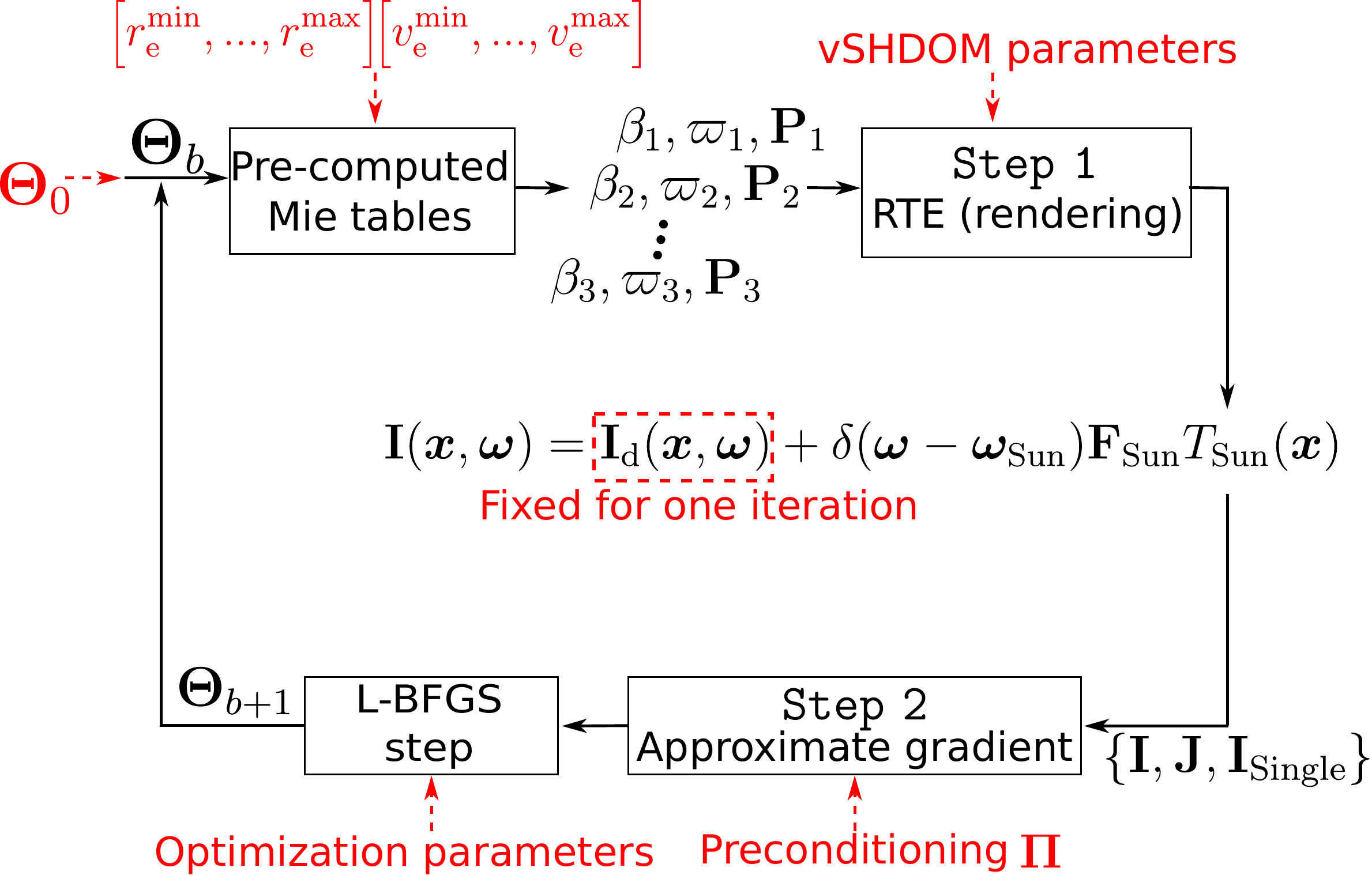}
		\caption{A block diagram of the iterative algorithm. Red marks hyper-parameter. Numerical parameters of vSHDOM and L-BFGS are summarized in the .}
		\label{fig:block_diagram}
	\end{figure}

	We formulate an iterative algorithm which alternates between two steps (See the diagram in Fig.~\ref{fig:block_diagram}). 
	Starting with an initial guess, ${\bf \Theta}_0$, {\tt Step 1} uses vSHDOM to compute the forward (recursive) RT equations. This renders synthetic images according to the multi-view geometry, spectral bands and spatial samples of the cameras. 
	Keeping ${\bf I}_{\rm d}$ fixed, {\tt Step 2} efficiently computes an approximate gradient with respect to ${\bf \Theta}$.
	The approximate gradient is fed into an L-BFGS step to update the current estimate ${\bf \Theta}_b$. 
	\\
	
	\noindent {\tt Step 1: RTE Forward Model}
	
	\noindent The first step in the estimation approach is running the forward model in Eqs.~[\ref{eq:rte_integral}-\ref{eq:J}] using a numerical RTE solver. This requires transforming microphysical to optical properties at every voxel ($g$) and spectral band ($\lambda$):
	\begin{equation}
	\label{eq:transform}
	{\rm LWC}[g],~ r_{\rm e}[g],~ v_{\rm e}[g] ~ \longrightarrow ~ \beta_{\lambda}[g], ~\varpi_{\lambda}[g], ~{\bf P}_{\lambda}[g].
	\end{equation}
	Implementing \eqref{eq:transform} using Eqs.~[\ref{eq:bulk_cs}-\ref{eq:bulk_p}] during each optimization iteration can be time-consuming. 
	Therefore, define grids \mbox{$r_{\rm e} {\in} \left[ r_{\rm e}^{\rm min},...,r_{\rm e}^{\rm max}\right]$} and \mbox{$v_{\rm e} {\in}\left[ v_{\rm e}^{\rm min},...,v_{\rm e}^{\rm max} \right]$}, for which lookup tables $\tilde{{\bf \sigma}}_{\lambda}\left(r_{\rm e}, v_{\rm e}\right),~ \varpi_{\lambda}\left(r_{\rm e}, v_{\rm e}\right),~ {\bf P}_{\lambda}\left(r_{\rm e}, v_{\rm e}\right)$ are pre-computed.
	With these pre-computed tables and $\{ {\rm LWC}[g], r_{\rm e}[g], v_{\rm e}[g]\}$, vSHDOM~\cite{doicu2013multi} renders the Stokes vector at each 3D voxel and direction. 
	This is the forward modeling procedure.
	The result is the set of fields ${\bf I}\xw, {\bf I}_{\rm d}\xw, {\bf J}\xw,
	{\bf I}_{\rm Single}\left({\bm x}_{\rm Sun} {\rightarrow} {\bm x}' {\rightarrow} {\bm x}_k \right)$. 
	\\
	
	\noindent {\tt Step 2: Approximate Jacobian Computation}
	
	\noindent The forward vRTE model in \eqref{eq:forward_stokes} depends on optical properties $\left(\beta, \varpi, {\bf P}\right)$, which themselves depend on the sought microphysics. 
	The Jacobian at voxel $g$ is expressed by applying the chain-rule to \eqref{eq:forward_stokes}. 
	For example, the derivative with respect to the effective radius is
	\begin{equation}
	\frac{\partial {\bf I}[k]}{\partial r_{\rm e}[g]} = \frac{\partial {\bf I}[k]}{\partial \beta[g]}\frac{\partial \beta[g]}{\partial r_{\rm e}[g]} + \frac{\partial {\bf I}[k]}{\partial \varpi[g]}\frac{\partial \varpi[g]}{\partial r_{\rm e}[g]} + \frac{\partial {\bf I}[k]}{\partial {\bf P}[g]}\frac{\partial {\bf P}[g]}{\partial r_{\rm e}[g]}.
	\label{eq:chain_rule}
	\end{equation}
	Analogously, replacing $r_{\rm e}$ in \eqref{eq:chain_rule} with ${\rm LWC}$ or $v_{\rm e}$ yields the respective microphysical derivatives. 
	We proceed by expressing the derivatives $\partial \{\beta, \varpi, {\bf P} \} / \partial \{{\rm LWC}, r_{\rm e}, v_{\rm e} \}$. 
	Afterwards,
	we expand and combine the derivatives $\partial {\bf I} / \partial \{\beta, \varpi, {\bf P} \}$ to express \eqref{eq:chain_rule}.
	
	For each voxel, the derivatives of $\left(\beta, \varpi, {\bf P}\right)$ with respect to the microphysics are calculated using pre-computed tables
	\begin{eqnarray}
	\frac{\partial \beta}{\partial \rm LWC}  = \tilde{{\bf \sigma}}(r_{\rm e},v_{\rm e}), \quad \! \frac{\partial  \beta}{\partial r_{\rm e}} = \frac{\tilde{{\bf \sigma}} (r_{\rm e} {+} \varepsilon_{r_{\rm e}}, v_{\rm e}) {-} \tilde{{\bf \sigma}} (r_{\rm e},v_{\rm e})}{\varepsilon_{r_{\rm e}}}, ~\\
	\frac{\partial \varpi}{\partial \rm LWC} = 0, \quad \quad \quad \quad \frac{\partial \varpi}{\partial r_{\rm e}}  = \frac{\varpi (r_{\rm e} {+} \varepsilon_{r_{\rm e}}, v_{\rm e}) {-} \varpi (r_{\rm e}, v_{\rm e})}{\varepsilon_{r_{\rm e}}}, \\
	\frac{\partial {\bf P}}{\partial \rm LWC} = 0,  \quad \quad \quad \quad \frac{\partial {\bf P} }{\partial r_{\rm e}} = \frac{{\bf P} (r_{\rm e} {+} \varepsilon_{r_{\rm e}}, v_{\rm e}) {-} {\bf P} (r_{\rm e},v_{\rm e})}{\varepsilon_{r_{\rm e}}}~,
	\label{eq:optical_derivatives}
	\end{eqnarray}
	where $v_{\rm e}$ derivatives are computed analogously to the $r_{\rm e}$ derivatives. 
	Using the shorthand notation \mbox{$\partial_g {\equiv} \{ \frac{\partial}{\partial {\rm LWC}[g]},~\frac{\partial}{\partial r_{\rm e}[g]},~\frac{\partial}{\partial v_{\rm e}[g]} \}$}, the overall Jacobian is given by a sum of terms
	\begin{equation}
	\partial_g {\bf I}[k]  =  A_1 + A_2 + A_3 + A_4 + A_5 + A_6.
	\label{eq:voxel_derivative}
	\end{equation}
	The full expression for each term in Eq.~[\ref{eq:voxel_derivative}]  is given in the {\it Appendix}. For example, 
	\begin{equation}
	A_1 = -  \ell_g \left({\bm x}_0 {\rightarrow} {\bm x}_k \right){\bf I}\left({\bm x}_0, {\bm \omega}_k \right) T\left({\bm x}_0 {\rightarrow} {\bm x}_k\right) \left[ \partial_g\beta \right]. \label{eq:A1}
	\end{equation}
	
	Let us focus on the term 
	\begin{eqnarray}
	A_4 &=& \int_{{\bm x}_0}^{{\bm x}_k} \bigg\{\frac{\varpi({\bm x}')}{4\pi}  \int_{4\pi}
	{\bf P}\left({\bm x}',{\bm \omega}_k{\cdot}{\bm \omega'}\right)  \big[ \partial_g {\bf I} \left({\bm x}',{\bm \omega'}\right)\big] {\rm d}{\bm \omega}' \bigg\}  
	\beta({\bm x}') T\left({\bm x}'  {\rightarrow} {\bm x}_k\right) {\rm d}{\bm x}'. \label{eq:A4}
	\end{eqnarray} 
	This Jacobian term stands out, because it is only term which requires computing the derivative of ${\bf I}$. This derivative is computationally expensive because ${\bf I}$ is computed recursively through the RTE [Eqs.~\ref{eq:rte_integral}-\ref{eq:J}]. 
	In principle, a change in the microphysics of one voxel can recursively affect the radiance at every other voxel. 
	We decompose $\partial_g {\bf I}$ using the diffuse-direct decomposition of \eqref{eq:I_seperation}
	\begin{eqnarray}
	\partial_g {\bf I} \left({\bm x}',{\bm \omega'}\right) = \partial_g {\bf I}_{\rm d} \left({\bm x}',{\bm \omega'}\right) 
	+  
	\delta \left({\bm \omega}' {-} {\bm \omega}_{\rm Sun} \right){\bf F}_{\rm Sun}  \big[ \partial_g T\left({\bm x}_{\rm Sun} {\rightarrow} {\bm x}'\right) \big]. \label{eq:A4_separation}
	\end{eqnarray}
	At the core our approach for computational efficiency is the assumption that the diffuse light ${\bf I}_{\rm d}$ is less sensitive to slight changes in the microphysical properties of any single voxel $g$. 
	Rather, ${\bf I}_{\rm d}$ is impacted mainly by bulk changes to the over-all volume. 
	Thus, we approximate \eqref{eq:A4} by keeping ${\bf I}_{\rm d}$ independent of ${\bm \Theta}$ for a single iteration of the gradient computation, i.e.,
	\begin{equation}
	\partial_g {\bf I}_{\rm d} {\approx} 0.
	\label{eq:approx}
	\end{equation}
	This bypasses the complexity of recursively computing $\partial_g {\bf I}_{\rm d}$. 
	
	It is important to note that at every iteration, the 
	Jacobian ${\bf \nabla}_{\bf \Theta} {\bf I}[k]$
	still {\bf is} impacted by ${\bf I}_{\rm d}$. This is because ${\bf I}_{\rm d}$ affects 
	${\bf I}$ through Eq.~[\ref{eq:I_seperation}], and  ${\bf I}$ appears in the terms $A_1,\ldots, A_6$. As the estimated medium properties evolve through iterations, so does ${\bf I}_{\rm d}$
	(in  {\tt Step 1}, above).
	We just assume during {\tt Step 2} that $\partial_g {\bf I}_{\rm d}$ is negligible compared to other terms in Eq.~[\ref{eq:voxel_derivative}]. 
	
	Contrary to ${\bf I}_{\rm d}$, the single-scattered component is highly sensitive to changes in the micro-physical properties of a single voxel. We therefore include an exact treatment of single-scattering in the gradient computation (in the {\it Appendix}).
	This is the essence of our numerical optimization approach. It enables tackling multiple-scattering tomography, in practice. 
	Simulation results presented in the following section rely on additional numerical considerations (e.g., initialization, preconditioning, convergence criteria), which are all described in the accompanying {\it Appendix}.

	\section{Simulations}
	\label{sec:simulations}
	
	As mentioned, real data of simultaneous spaceborne multi-angular polarimetric images of clouds does not yet exist, but a mission to supply this data is in the works. Therefore, we use careful simulations to test the approach. We simulate an atmosphere with molecular Rayleigh scattering and liquid water clouds. Rayleigh scattering is taken from the AFGL database~\cite{anderson1986afgl} for a summer mid-latitude atmosphere.  
	Mie tables are pre-computed for $r_{\rm e} \in [4, 25]\,\mu$m and $v_{\rm e} = 0.1$ with $N_{r_{\rm e}} = 100$. 
	The surface is Lambertian with a water-like albedo of $0.05$. 
	For realistic complexity, a Large Eddy Simulation (LES) model~\cite{matheou2014large} was used to generate a cloud field. Each voxel is of size \mbox{$20{\times} 20{\times} 40\,{\rm m}^3$}.
	The LES outputs~\cite{matheou2014large} are clouds with 3D variable ${\rm LWC}$ and 1D (vertically) variable $r_{\rm e}$. A typical value~\cite{yau1996short} of $v_{\rm e}=0.1$ was chosen. Consequently, the present recovery demonstrations recover ${\rm LWC}$ and $r_{\rm e}$ on their respective native LES grid. On the other hand, $v_{\rm e} = 0.1$ is excluded from the unknowns.
	
\begin{figure}[t]
		\centering \includegraphics[width=0.6\linewidth]{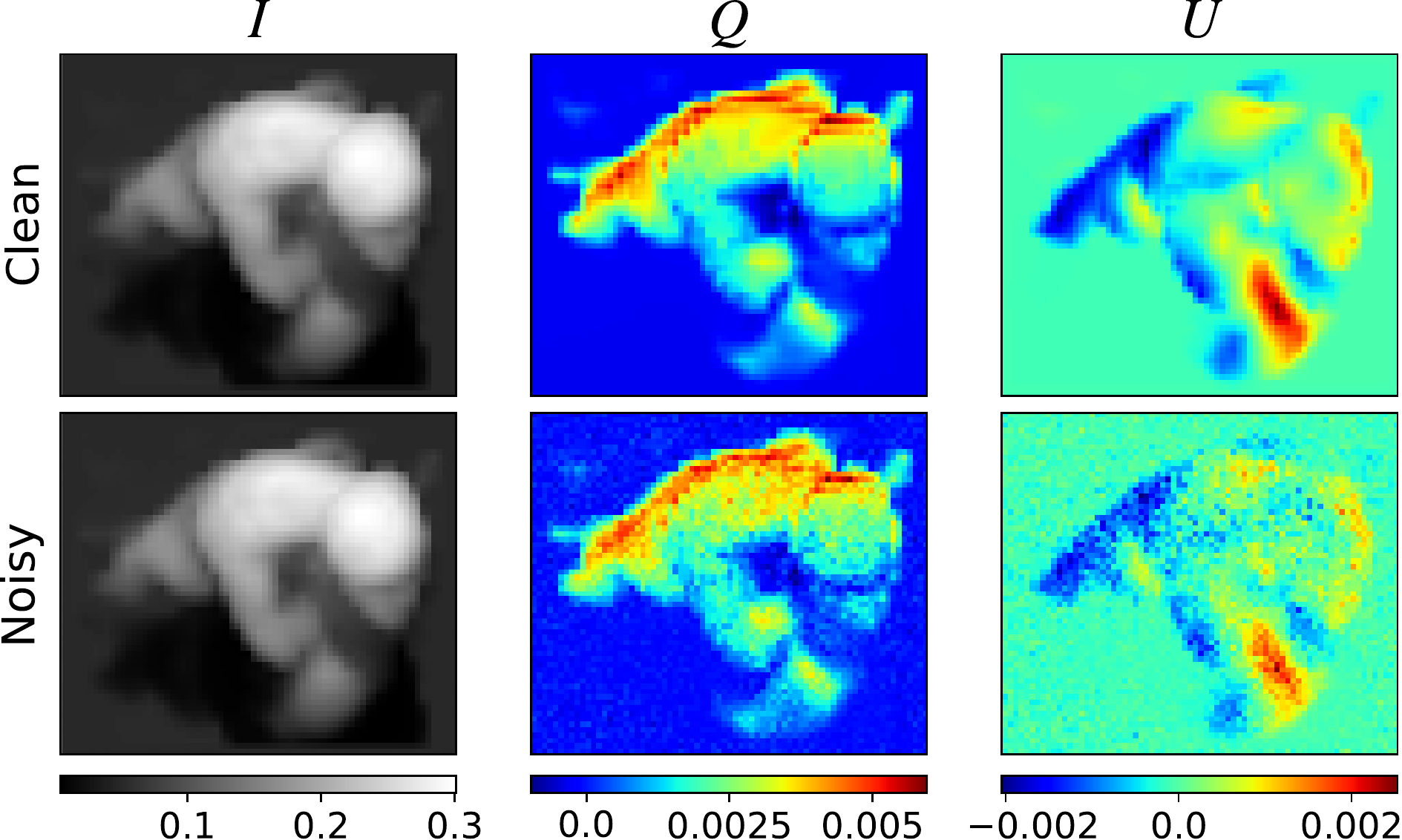}
		\caption{{\tt Scene A} synthesized Stokes image using vSHDOM, before and after the application of a realistic AirMSPI noise model. We show here the {\em Bidrectional Reflectance Factor} (BRF) of the nadir view at $\lambda {=}0.67\mu {\rm m}$.}
		\label{fig:noise}
	\end{figure}

\begin{figure}[t]
		\centering \includegraphics[width=0.8\linewidth]{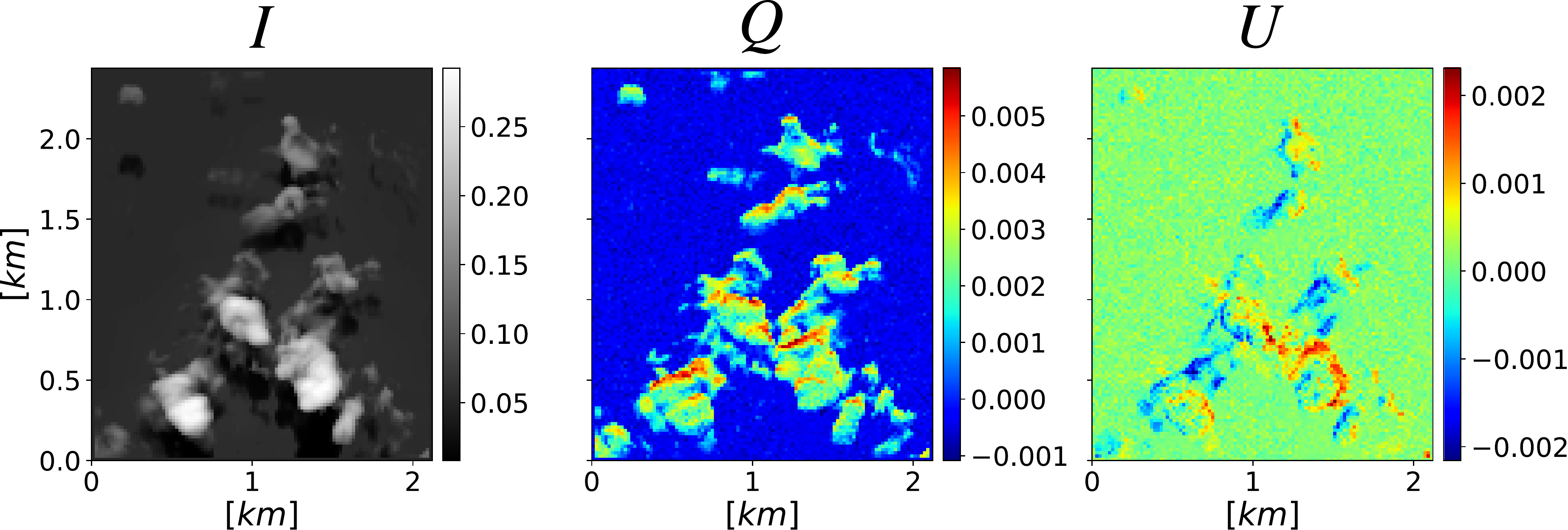}
		\caption{{\tt Scene B} synthesized Stokes using vSHDOM. We show here the BRF of the nadir view at $\lambda {=} 0.67\mu {\rm m}$.} 
		\label{fig:nadir_view2}
	\end{figure}

	From the generated cloud field, two isolated cloudy regions are taken for reconstruction:
	\begin{enumerate}[leftmargin=0.7cm,itemsep=-1pt,topsep=2pt]
		\item {\tt Scene A:} An atmospheric domain of dimensions $0.64{\times} 0.72{\times} 20~{\rm km}^3$ with an isolated cloud (see synthetic AirMSPI nadir view in Fig.~\ref{fig:noise}). 
		\item {\tt Scene B:}  An atmospheric domain of dimensions $2.42{\times} 2.1{\times} 8~{\rm km}^3$ with several clouds of varying optical thickness (see synthetic AirMSPI nadir view in Fig.~\ref{fig:nadir_view2}).
	\end{enumerate}
	Synthetic measurements rendered with the spatial resolution and angular sampling of AirMSPI~\cite{diner2013airborne}, namely, $10$~m pixels and 9 viewing angles: ${\pm}70.5^\circ$, ${\pm}60^\circ$, ${\pm}45.6^\circ$, ${\pm}26.1^\circ$, and $0^\circ$ from zenith, where ${\pm}$ indicates fore- and aft-views along the northbound flight path. 
	Solar zenith angle is $15^\circ$ from nadir in the measurement plane, i.e., $0^\circ$ solar azimuth. 
	We simulate measurements at AirMSPI's three polarized spectral bands: $\lambda = \left[ 0.47, 0.66, 0.865 \right]~\mu$m. The bandwidths are narrow enough ($\approx$46~nm) to render images using a single representative wavelength per band.
	
	Single scattering albedos for these wavelengths are all within $10^{-4}$ of unity. 
	In other words, and in sharp contrast with the operational Nakajima--King \cite{nakajima1990determination} bi-spectral non-tomographic retrieval, absorption by droplets plays no role in this demonstration of tomography of cloud microphysics.
	The measurements are synthesized with realistic noise, according to the AirMSPI data acquisition model (see {\it Appendix}).
	
\begin{figure}[t]
		\centering 
		\includegraphics[width=0.35\linewidth]{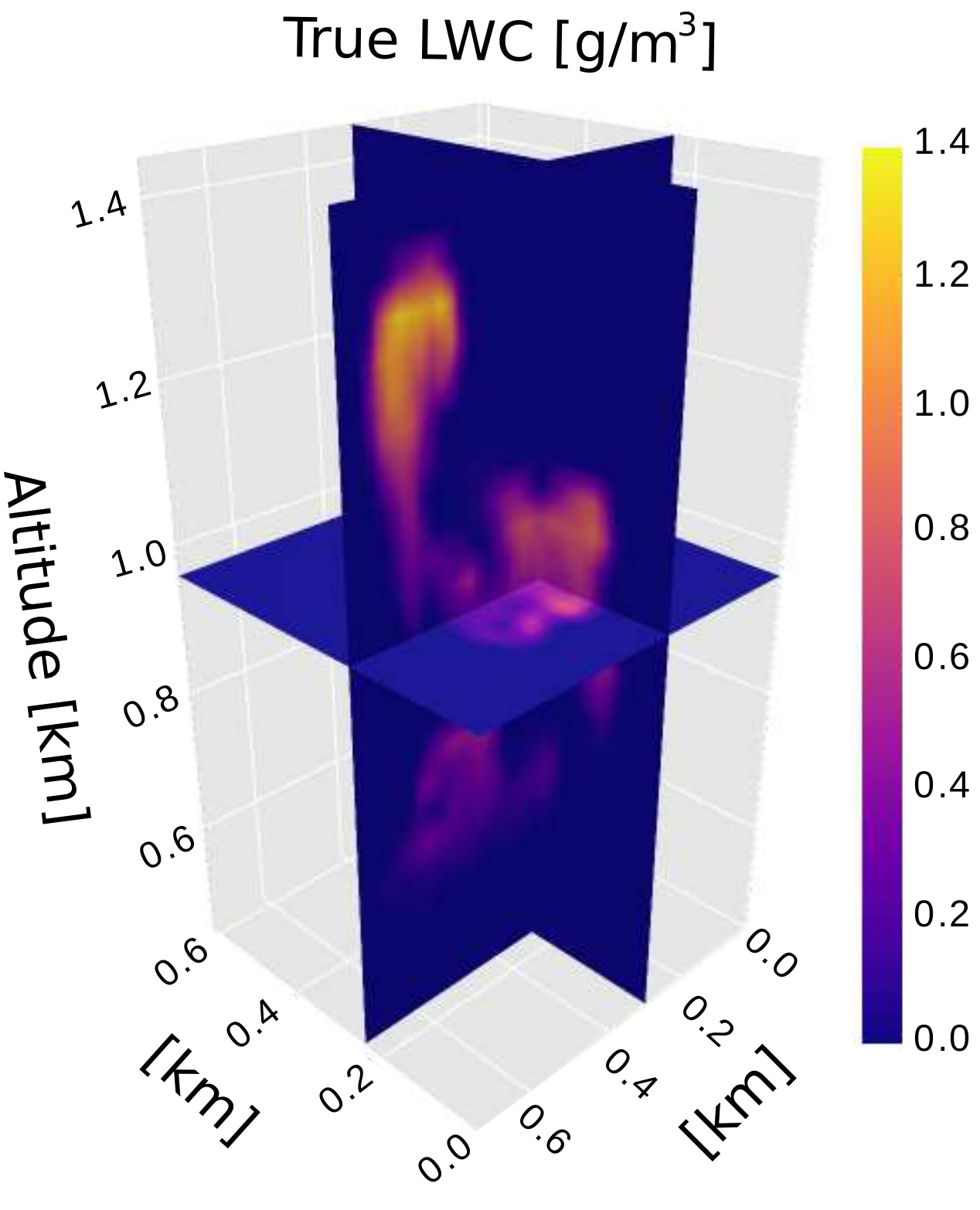}~~~~~~~~~~~~~~~~~~
		\includegraphics[width=0.35\linewidth]{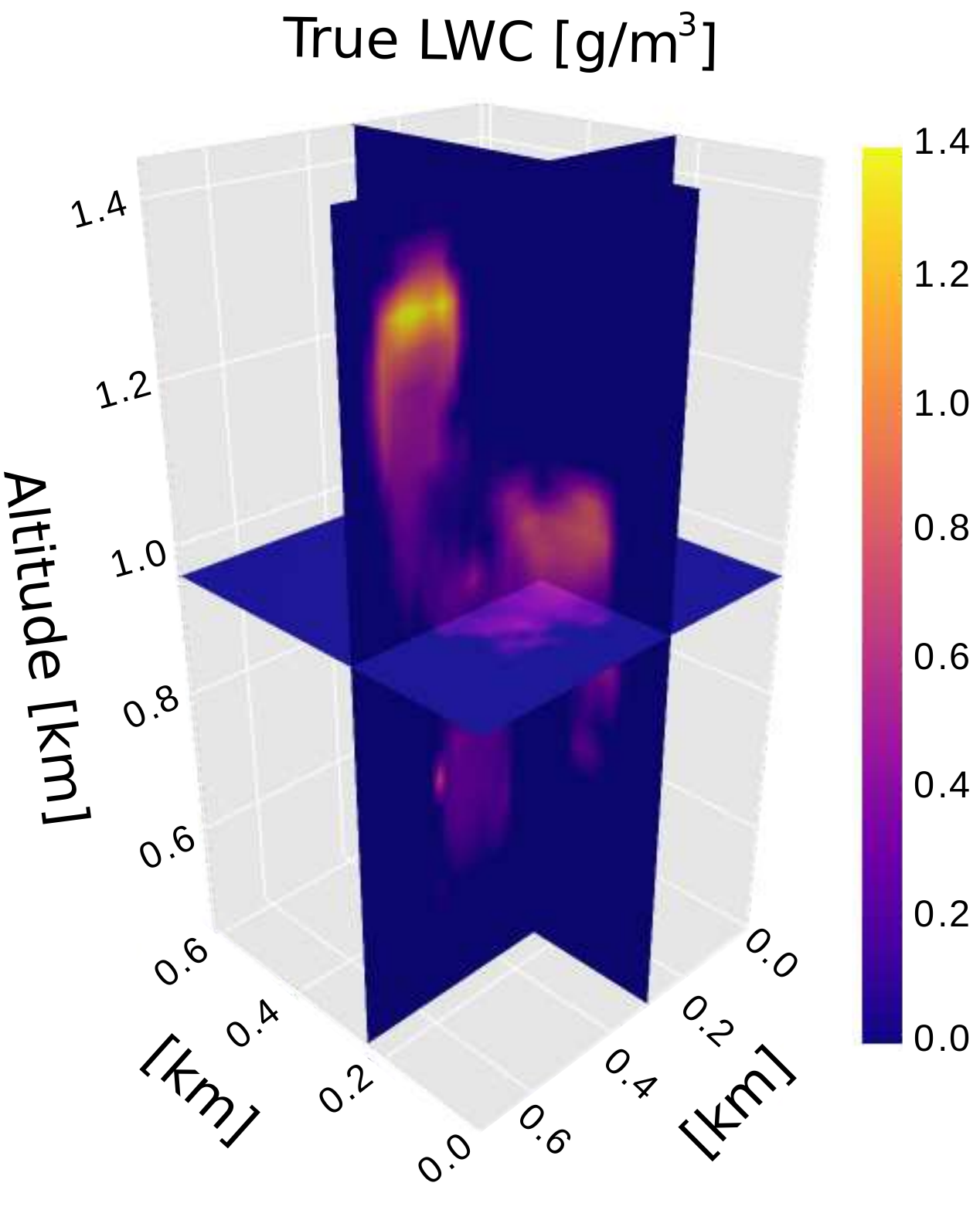}
		\caption{{\tt Scene A} recovery results. [Left] Slices of the true cloud generated by LES. [Right] Slices of the cloud estimated tomographically using AirMSPI polarized bands.}
		\label{fig:lwc1_true}
	\end{figure}

\begin{figure}[t]
		\centering \includegraphics[width=0.35\linewidth]{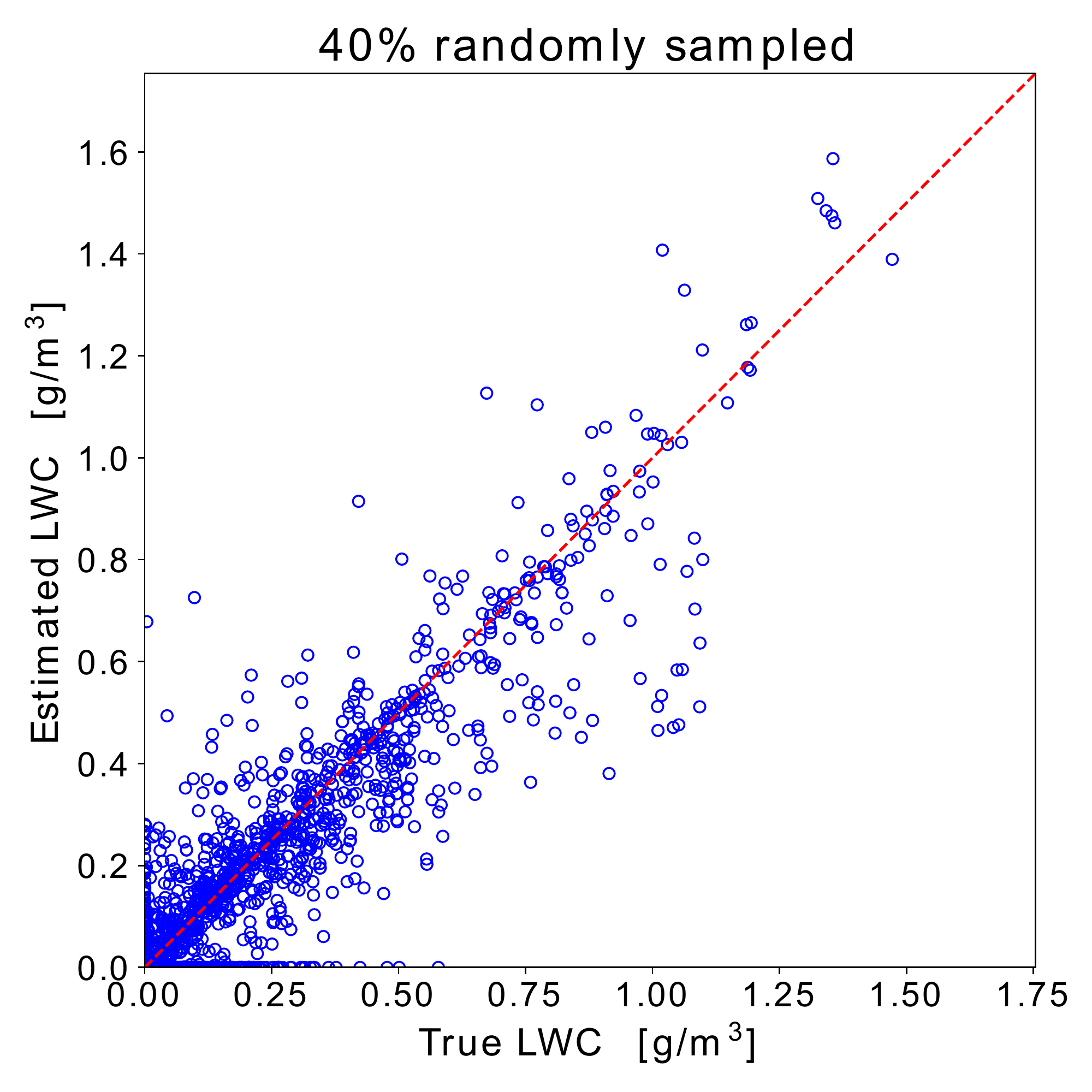}~~~~~~~~~~~~~~~~~~
		\includegraphics[width=0.35\linewidth]{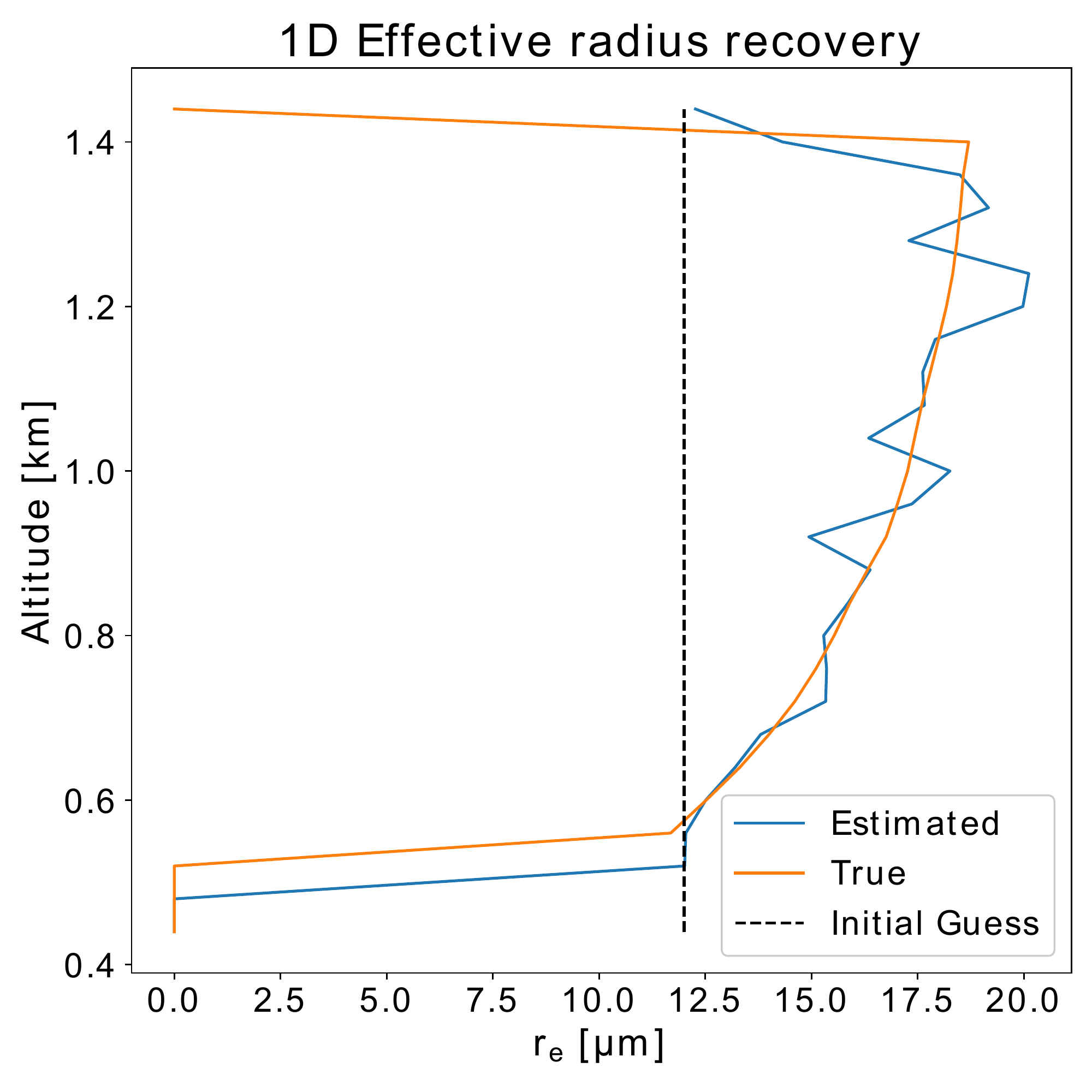}
		\caption{{\tt Scene A} recovery results. [Left] Scatter plot of estimated vs. true LWC. The correlation coefficient is 0.94. [Right] recovery results of the 1D effective radius.}
		\caption{{\tt Scene A} }
		\label{fig:scatter1}
	\end{figure}

	Qualitative volumetric results of the recovered LWC for {\tt Scene A} are shown in Fig.~\ref{fig:lwc1_true}.
	Scatter plot of the recovered LWC and the recovery results of $r_{\rm e}$ for {\tt Scene A} are given in Fig.~\ref{fig:scatter1}. 
	Analogous plots for {\tt Scene B} recovery results are given in the {\it Appendix}. 
	
	For quantitative assessment of the recovery, we use local mean error $\epsilon$, and global bias measures~\cite{aides2013multi} $\vartheta$:
	\begin{eqnarray}
	\epsilon_{\rm LWC} = \frac{\|\hat{\rm  LWC} {-} {\rm  LWC}\|_1}{\|{\rm  LWC}\|_1},~~ \vartheta_{\rm LWC} = \frac{\|\hat{\rm  LWC}\|_1 {-} \|{\rm  LWC}\|_1}{\|{\rm  LWC}\|_1}, ~~
	\epsilon_{r_{\rm e}} = \frac{\|\hat{r_{\rm e}} - r_{\rm e}\|_1 }{\|r_{\rm e}\|_1}.
	\end{eqnarray}
	The quantitative error measures upon convergence for the two scenes are:\\
	\-\hspace{0.1cm} {\tt Scene A:} \mbox{$\epsilon_{r_{\rm e}}{\approx} 11 \%,~\epsilon_{\rm LWC} {\approx} 30 \%,~\vartheta_{\rm LWC} {\approx} -4 \%$}, \\
	\-\hspace{0.1cm} {\tt Scene B:} \mbox{$\epsilon_{r_{\rm e}}{\approx} 13 \%,~\epsilon_{\rm LWC} {\approx} 29 \%,~\vartheta_{\rm LWC} {\approx} -5 \%$}. \\
	Using a 2.50~GHz CPU, the recovery run-time of cloud properties in {\tt Scenes A,B} was ${\sim} 13$ hours and ${\sim} 10$ days, respectively. 
	
	Multi-angular tomographic retrieval enables vertical resolution of the droplet effective radius. By contrast, a homogeneous droplet radius is typically retrieved by mono-angular observations fitted to a plane-parallel homogeneous cloud model. The retrieval errors of  droplet radii in the demonstrations above
	are significantly smaller  than retrieval errors of a homogeneous droplet radius. The latter can easily exceed 50\% in similar conditions to our study i.e, shallow cumuli and illumination conditions (see e.g.~\cite{Seethala2012EvaluatingTS}).
	
	\section{Summary \& Outlook}
	\label{sec:conclusion}
	We derive tomography of cloud microphysics based on multi-view/multi-spectral polarimetric measurements of scattered sunlight. This novel type of tomography uses, for the first time, 3D polarized RT as the image formation model. 
	We define a model-fitting error function and compute approximate gradients of this function to make the recovery tractable. Demonstration are done on synthetic 3D clouds, based on a Large Eddy Simulation with the effective radius assumed to vary only vertically. 
	
	Future work will address the extent to which polarimetric measurements penetrate the cloud and the relation between $r_{\rm e}$ in the {\em outer shell} and $r_{\rm e}$ in the cloud {\em core}, as defined by Forster et al.~\cite{forster2019cloud}. 
	Furthermore, we will relax the fixed $v_{\rm e}$ assumption that was used in the demonstrations, and thus assess full microphysical retrieval capabilities of polarization measurements. 
	A thorough discussion on these assumptions and their applicability to real-world clouds is given in the {\it Appendix}. Moreover, future  plans include experimental demonstration and use, while the CloudCT formation orbits.
	
	Lastly, we note that our atmospheric tomography approach herein can be adapted to aerosols, including dense plumes of wild fire smoke, volcanic ash, and dust. 
	Research is ongoing~\cite{forster2019cloud} about such adaptation 
	for satellite data as can be obtained from the multi-view imaging from MISR on Terra and a SWIR view from the collocated MODIS, as well as in the planned CloudCT~\cite{cloudct}.
	
	{\small
		\section*{Acknowledgements}
		\label{sec:acknowledgements}
		We thank I.~Koren, D.~Rosenfeld, A.~Aides, D.~Diner, L.~Di~Girolamo, and G.~Matheou for support and fruitful discussions.
		We acknowledge F.~Evans and A.~Doicu for the online vSHDOM code. 
		The authors are grateful to the US-Israel Binational Science Foundation (BSF grant 2016325) for continuously facilitating our international collaboration. 
		Aviad Levis work was partially supported by the Zuckerman Foundation.
		Yoav Schechner is a Landau Fellow supported by the Taub Foundation. 
		His work was conducted in the Ollendorff Minerva Center (BMBF). 
		Anthony Davis' work was carried out at JPL/Caltech, supported by NASA's SMD/ESD/(RST+TASNPP) and ESTO/AIST programs. 
		Support for Jesse Loveridge's work from JPL under contract \#147871 is gratefully acknowledged. 
		This project has received funding from the European Research Council (ERC) under the European Union’s Horizon 2020 research and innovation program (grant agreement No 810370: CloudCT).
	}
	
	\clearpage
	
	\bibliographystyle{pnas2009} 
	\bibliography{references}
	
	\newpage
	\appendix
	\appendixpage
	
	\section{Jacobian Derivation}
	In Eq.~[\ref{eq:voxel_derivative}] of the main text,  the Jacobian is written as a sum of six terms
	\begin{equation}
	\partial_g {\bf I}[k]  =  A_1 + A_2 + A_3 + A_4 + A_5 + A_6.
	\label{supeq:voxel_derivative}
	\end{equation}
	In this section we expand and describe each of these terms. Using Eqs.~[\ref{eq:T}] and [\ref{eq:numerical_integration}], the transmittance derivative is
	\begin{equation}
	\partial_g T\left({\bm x}_1 {\rightarrow} {\bm x}_2\right)= - T\left({\bm x}_1 {\rightarrow} {\bm x}_2\right) \ell_g \left({\bm x}_1 {\rightarrow} {\bm x}_2 \right)\partial_g\beta.
	\label{supeq:trans_grad}
	\end{equation}
	Then,
	\begin{eqnarray}
	A_1  &=&  -  \ell_g \left({\bm x}_0 {\rightarrow} {\bm x}_k \right){\bf I}\left({\bm x}_0, {\bm \omega}_k \right) T\left({\bm x}_0 {\rightarrow} {\bm x}_k\right) \left[ \partial_g\beta \right], \label{supeq:A1} \\
	A_2 &=&  \ell_g \left({\bm x}_0 {\rightarrow} {\bm x}_k \right)\int\limits^{{\bm x}_k}_{{\bm x}_0}  \bigg[\frac{\partial_g \varpi} {4\pi} \int\limits_{4\pi}
	{\bf P}\left({\bm x}',{\bm \omega}_k{\cdot}{\bm \omega'}\right) {\bf I} \left({\bm x}',{\bm \omega'}\right)   {\rm d}{\bm \omega}' \bigg] \beta({\bm x}') T\left({\bm x}'  {\rightarrow} {\bm x}_k\right) {\rm d}{\bm x}', \\
	A_3 &=& \ell_g \left({\bm x}_0 {\rightarrow} {\bm x}_k \right)\int\limits^{{\bm x}_k}_{{\bm x}_0} \bigg\{\frac{\varpi({\bm x}')}{4\pi}  \int\limits_{4\pi}
	\big[ \partial_g  {\bf P}\left({\bm x}',{\bm \omega}_k{\cdot}{\bm \omega'}\right) \big] {\bf I} \left({\bm x}',{\bm \omega'}\right) {\rm d}{\bm \omega}' \bigg\} \beta({\bm x}') T\left({\bm x}'  {\rightarrow} {\bm x}_k\right) {\rm d}{\bm x}', \\
	A_4 &=& \int\limits^{{\bm x}_k}_{{\bm x}_0} \bigg\{\frac{\varpi({\bm x}')}{4\pi}  \int\limits_{4\pi}
	{\bf P}\left({\bm x}',{\bm \omega}_k{\cdot}{\bm \omega'}\right)  \big[ \partial_g {\bf I} \left({\bm x}',{\bm \omega'}\right)\big] {\rm d}{\bm \omega}' \bigg\} \beta({\bm x}') T\left({\bm x}'  {\rightarrow} {\bm x}_k\right) {\rm d}{\bm x}', \label{supeq:A4}\\
	A_5 &=& \ell_g \left({\bm x}_0 {\rightarrow} {\bm x}_k \right) \left[ \partial_g\beta\right] \int\limits^{{\bm x}_k}_{{\bm x}_0}  {\bf J}\left({\bm x}', {\bm \omega}_k \right) T\left({\bm x}'  {\rightarrow} {\bm x}_k\right) {\rm d}{\bm x}', \\
	A_6 &=& -\ell_g \left({\bm x}_0 {\rightarrow} {\bm x}_k \right) \left[ \partial_g\beta \right] \int\limits^{{\bm x}_k}_{{\bm x}_0}  {\bf J}\left({\bm x}', {\bm \omega}_k \right) \beta({\bm x}') T\left({\bm x}'  {\rightarrow} {\bm x}_k\right) {\rm d}{\bm x}'.
	\label{supeq:A6}
	\end{eqnarray}
	Note that ${\bf I} \xw$ and ${\bf J} \xw$ are computed in {\tt Step 1} and are therefor ready for use when computing $A_1,A_2,A_3,A_5$ and $A_6$. Furthermore, $\ell_g \left({\bm x}_0 {\rightarrow} {\bm x}_k \right){=}0$ for any voxel that is not on the LOS of pixel $k$. Therefore, the terms $A_1,A_2,A_3,A_5,A_6$ are computed using a single path tracing ${\bm x}_k {\rightarrow} {\bm x}_0$.
	
	We now give special attention to $A_4$ in Eq.~[\ref{supeq:A4}]. Using the diffuse-direct decomposition of \eqref{eq:I_seperation}, we decompose  \eqref{supeq:A4} as
	\begin{eqnarray}
	A_4 & =& \int\limits^{{\bm x}_k}_{{\bm x}_0} \bigg\{\frac{\varpi({\bm x}')}{4\pi}  \int\limits_{4\pi}
	{\bf P}\left({\bm x}',{\bm \omega}_k{\cdot}{\bm \omega'}\right)  \big[ \partial_g {\bf I}_{\rm d} \left({\bm x}',{\bm \omega'}\right) \big] {\rm d}{\bm \omega}' \bigg\} \beta({\bm x}') T\left({\bm x}'  {\rightarrow} {\bm x}_k\right) {\rm d}{\bm x}' \nonumber \\
	& + & 
	\int\limits^{{\bm x}_k}_{{\bm x}_0} \bigg\{\frac{\varpi({\bm x}')}{4\pi}  \int\limits_{4\pi}
	{\bf P}\left({\bm x}',{\bm \omega}_k{\cdot}{\bm \omega'}\right) \delta \left({\bm \omega}' {-} {\bm \omega}_{\rm Sun} \right){\bf F}_{\rm Sun}   \big[ \partial_g T\left({\bm x}_{\rm Sun} {\rightarrow} {\bm x}'\right) \big] {\rm d}{\bm \omega}' \bigg\} \beta({\bm x}') T\left({\bm x}'  {\rightarrow} {\bm x}_k\right) {\rm d}{\bm x}'.
	\label{supeq:A4_separation}
	\end{eqnarray}
	The first term in \eqref{supeq:A4_separation} is based on $\partial_g {\bf I}_{\rm d}$, ie., a derivative of the diffuse (high order scattering) component. Herein lies a recursive complexity.
	In principle, a differential change in the microphysics of one voxel can recursively affect the radiance at every other voxel, and this affects all the pixels. To make calculations numerically efficient,  we approximate \eqref{supeq:A4_separation}. The approximation assumes that relative to other components in the Jacobian, ${\bf I}_{\rm d}$ is less sensitive to a differential changes in the microphysical properties at voxel $g$.  Thus, \eqref{supeq:A4_separation} is approximated by keeping ${\bf I}_{\rm d}$ independent of ${\bm \Theta}$ for a single iteration of the gradient computation, i.e, 
	\begin{equation}
	\partial_g {\bf I}_{\rm d} {\approx} 0 \;.
	\label{supeq:dId}
	\end{equation}
	
	The second term in \eqref{supeq:A4_separation} is based on differentiation of the direct component. This is straight-forward to compute using \eqref{supeq:trans_grad}. Consequently, using Eq.~[\ref{supeq:dId}] and the definition of ${\bf I}_{\rm Single}\left({\bm x}_{\rm Sun} {\rightarrow} {\bm x}' {\rightarrow} {\bm x}_k \right)$ in \eqref{eq:single_scatter_x}, the term $A_4$ in \eqref{supeq:A4_separation} is approximated by
	\begin{equation}
	A_4 \approx \tilde{A_4} =  \left[ \partial_g\beta\right] \int\limits^{{\bm x}_k}_{{\bm x}_0} \ell_g    \left({\bm x}_{\rm Sun} {\rightarrow} {\bm x}' \right){\bf I}_{\rm Single}\left({\bm x}_{\rm Sun} {\rightarrow} {\bm x}' {\rightarrow} {\bm x}_k \right){\rm d}{\bm x}'.
	\label{supeq:direct_derivative}
	\end{equation}
	The term $\ell_g \left({\bm x}_{\rm Sun} {\rightarrow} {\bm x}' \right)$ in \eqref{supeq:direct_derivative} contributes to voxels {\em outside} of the LOS. The integral in $\tilde{A_4}$ is computed with a {\em broken-ray}~\cite{florescu2011inversion} path ${\bm x}_k {\rightarrow} {\bm x}' {\rightarrow} {\bm x}_{\rm Sun}$, as illustrated in Fig.~\ref{fig:domain}.

	Using Eqs.~[\ref{eq:rte_integral},\ref{supeq:trans_grad},\ref{supeq:A1},\ref{supeq:A6}], $A_1$ and $A_6$ are combined to 
	\begin{equation}
	A_{1,6} = A_1 + A_6 = - \left[ \partial_g\beta\right] {\bf I} \left({\bm x}_g, {\bm \omega}_k \right).
	\label{supeq:relation}
	\end{equation}
	Overall, in our iterative procedure, we approximate the Jacobian in \eqref{eq:voxel_derivative}  by
	\begin{eqnarray}
	\partial_g {\bf I}[k]  = A_{1,6} + A_2 + A_3 + \tilde{A_4} + A_5.
	\label{supeq:approximate_jacobian}
	\end{eqnarray}
	Equations~[\ref{supeq:A1}]--[\ref{supeq:direct_derivative}] formulate the Jacobian in terms of a voxel grid (zero-order interpolation). However, in practice we use a trilinear interpolation kernel $K$ in \eqref{eq:interpolation_kernel}, consistent with vSHDOM internal interpolation~\cite{evans1998spherical}.
	
	
	\section{Measurement Noise}
	The inverse problem defined in the main text is formulated in terms of measured Stokes vectors  [Eq.~\ref{eq:meas_vec}]. However, Stokes vectors are not measured directly. Rather, they are derived from intensity measurements taken through filters. The raw intensity measurements are noisy. Noise is dominated by Poisson photon noise, which is independent across different raw measurements. However, the estimation of Stokes components {\em from} independent intensity measurements yields noise which is correlated across the components of the Stokes vector, per-pixel. In this section, we describe the synthesis model we employ to generate realistic noise in simulations.
	Our synthesis is based on the AirMSPI~\cite{van2018calibration} sensor model. Furthermore, we derive the expression for ${\bf R}$, which we use in the recovery process (Eq.~[\ref{eq:minimization}] in the main text).
	
\begin{figure}
		\def\svgwidth{0.8\textwidth}
		\centering 
		
		\begingroup%
		\makeatletter%
		\providecommand\color[2][]{%
			\errmessage{(Inkscape) Color is used for the text in Inkscape, but the package 'color.sty' is not loaded}%
			\renewcommand\color[2][]{}%
		}%
		\providecommand\transparent[1]{%
			\errmessage{(Inkscape) Transparency is used (non-zero) for the text in Inkscape, but the package 'transparent.sty' is not loaded}%
			\renewcommand\transparent[1]{}%
		}%
		\providecommand\rotatebox[2]{#2}%
		\newcommand*\fsize{\dimexpr\f@size pt\relax}%
		\newcommand*\lineheight[1]{\fontsize{\fsize}{#1\fsize}\selectfont}%
		\ifx\svgwidth\undefined%
		\setlength{\unitlength}{508.44328633bp}%
		\ifx\svgscale\undefined%
		\relax%
		\else%
		\setlength{\unitlength}{\unitlength * \real{\svgscale}}%
		\fi%
		\else%
		\setlength{\unitlength}{\svgwidth}%
		\fi%
		\global\let\svgwidth\undefined%
		\global\let\svgscale\undefined%
		\makeatother%
		\begin{picture}(1,0.1809971)%
		\lineheight{1}%
		\setlength\tabcolsep{0pt}%
		\put(0,0){\includegraphics[width=\unitlength,page=1]{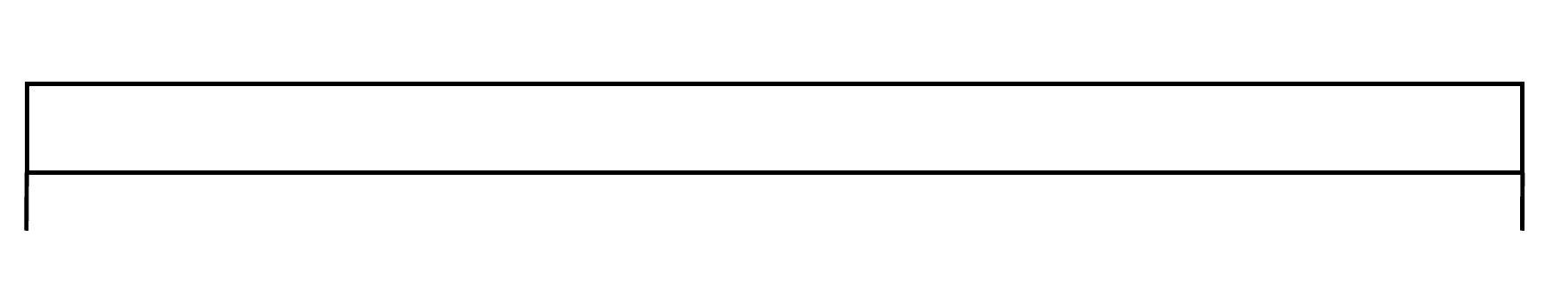}}%
		\put(-0.0025119,0.00700272){\color[rgb]{0,0,0}\makebox(0,0)[lt]{\lineheight{1.25}\smash{\begin{tabular}[t]{l}$-0.5$\end{tabular}}}}%
		\put(0,0){\includegraphics[width=\unitlength,page=2]{subframe.pdf}}%
		\put(0.48146543,0.15464435){\color[rgb]{0,0,0}\makebox(0,0)[lt]{\lineheight{1.25}\smash{\begin{tabular}[t]{l}$\psi_l$\end{tabular}}}}%
		\put(0,0){\includegraphics[width=\unitlength,page=3]{subframe.pdf}}%
		\put(0.03304296,0.15545671){\color[rgb]{0,0,0}\makebox(0,0)[lt]{\lineheight{1.25}\smash{\begin{tabular}[t]{l}$\psi_0$\end{tabular}}}}%
		\put(0.94477292,0.0058503){\color[rgb]{0,0,0}\makebox(0,0)[lt]{\lineheight{1.25}\smash{\begin{tabular}[t]{l}$0.5$\end{tabular}}}}%
		\put(0,0){\includegraphics[width=\unitlength,page=4]{subframe.pdf}}%
		\put(0.45242054,0.0128213){\color[rgb]{0,0,0}\makebox(0,0)[lt]{\lineheight{1.25}\smash{\begin{tabular}[t]{l}$\Delta \psi = 1/N_{\rm sub}$\end{tabular}}}}%
		\end{picture}%
		\endgroup%
		
		\caption{A normalized frame spans the interval $[-0.5,0.5]$, evenly divided into $N_{\rm sub}$ subframes.}
		\label{fig:subframe}
	\end{figure}

	AirMSPI measures a modulated intensity signal at $N_{\rm sub} {=} 23$ subframes. Define a normalized frame which spans the unitless integration time interval $\psi \in [-0.5, 0.5]$. Denote the temporal center and span of each subframe as $\psi_l$ and $\Delta \psi = 1/N_{\rm sub}$, respectively (Fig.~\ref{fig:subframe}). Based on the sensing process described in Ref.~\cite{van2018calibration}, define the following modulation function, whose parameters are given in Table~\ref{table:mspi_params}:
	\begin{equation}
	M[l] = J_0[\kappa(\psi_l)] + \frac{1}{3}\left(\frac{\pi \Delta \psi}{2} \right)^2\gamma_0^2(\lambda) \bigg\{ J_2[\kappa(\psi_l)] - \cos[2(\pi \psi_l - \eta)]J_0[\kappa(\psi_l)] \bigg\},
	\label{eq:Ml}
	\end{equation}
	with 
	\begin{equation}
	\kappa(\psi_l) = -2 \gamma_0(\lambda) \sin(\pi \psi_l - \eta)\sqrt{1+\cot^2(\pi \psi_l - \eta)}.
	\end{equation}
	\begin{table}
		\centering
		\caption{Modulation parameters~\cite{van2018calibration} used for synthesis of AirMSPI measurements.} 
		\label{table:mspi_params}
		\begin{tabular}{ccccccc} \toprule
			{$\gamma_0(470 {\rm nm})$} & {$\gamma_0(660 {\rm nm})$} & {$\gamma_0(865 {\rm nm})$} & {$\xi(470 {\rm nm})$} & {$\xi(660 {\rm nm})$} & {$\xi(865 {\rm nm})$} & {$\eta$}   \\ \midrule
			4.472 & 3.081 & 2.284 & 1.0 & 0.27 & 0.03 & 0.009 \\
		\end{tabular}
	\end{table}
	Here $J_0$, $J_2$ are the Bessel functions of the first kind of order 0 and 2, respectively. Denote by $\xi(\lambda)$ a wavelength-dependent ratio, which is drawn from quantum efficiencies and spectral bandwidths\footnote{For the exact calculation of the ratio see Eq.~[24] of \cite{van2018calibration}.} of each AirMSPI band (Table~\ref{table:mspi_params}). Using simulated Stokes vectors derived by vSHDOM, AirMSPI measurements are synthesised as passing through two polarization analyzing filters~\cite{van2018calibration}. 
	As defined in Eq.~[\ref{eq:forward_stokes}] in the main text, ${\bf I}[k]$ is the Stokes vector in pixel $k$. Correspondingly the intensity is $I[k]$, while $Q[k],U[k]$ are the polarized components.
	Measurements $l$ through the two filters of AirMSPI are modeled by
	\begin{eqnarray}
	\label{supeq:mspi_meas1}
	&d_0[l,k] = \xi(\lambda) \big(I[k] + M[l]Q[k] \big) \\
	&d_{45}[l,k] = \xi(\lambda) \big(I[k] + M[l]U[k] \big),
	\label{supeq:mspi_meas2}
	\end{eqnarray}
	where $M[l]$ and $\xi(\lambda)$ are given in Eq.~[\ref{eq:Ml}] and Table~\ref{table:mspi_params}, respectively.  The units of $d$ are Watts. Define ${\bf d}[k] = \left(d_0[1,k],...,d_0[N_{\rm sub},k], d_{45}[1,k],...,d_{45}[N_{\rm sub},k] \right)^\top$. In matrix from, the transformations by Eqs.~[\ref{supeq:mspi_meas1}-\ref{supeq:mspi_meas2}] are written using a single $46 {\times} 3$ modulation matrix ${\bf M}$ 
	\begin{equation}
	{\bf d}[k] = {\bf M} {\bf I}[k].
	\end{equation}
	
	Detection is by a camera which generates photo-electrons in each pixel well. The relation between $d_0[l,k]$ or $d_{45}[l,k]$ and the {\em expected} unit-less number of photo-electrons in the pixel is given by a gain $G$. The number of photo-electrons is random ({\em Poissonian}) around this expected value. The vector of simulated electron counts is thus synthesized by a Poisson process
	\begin{equation}
	{\bf e}[k] \sim Poisson \Big\{  {\rm round} \left( G  \cdot {\bf d}[k] \right) \Big\} = Poisson \Big\{  {\rm round} \left( G  \cdot {\bf M} {\bf I}[k] \right) \Big\}.
	\label{supeq:measurement_noise}
	\end{equation}
	The gain $G$ is chosen to let the maximum signal at each camera view (i.e. maximum over pixels, wavelengths and subframe measurements) reach the maximum {\em full-well depth} of $200,\!000$ electrons, consistent with AirMSPI specifications. \eqref{supeq:measurement_noise} synthesizes raw AirMSPI signals including noise (Fig.~\ref{fig:noise}).  The synthesized AirMSPI signals, including this noise, are now used as inputs to the calculation of measured Stokes vectors in each pixel and viewpoint. The vector of electron counts ${\bf e}[k]$ in each pixel $k$ is transformed into Stokes synthetic data [Eq.~\ref{eq:meas_vec}] using a $3 {\times} 46$ demodulation matrix ${\bf W}$
	\begin{equation}
	{\bf y}_{\bf I}[k] = ({\bf M}^\top {\bf M})^{-1}{\bf M}^\top{\bf e}[k] = {\bf W} {\bf e}[k].
	\label{supeq:y}
	\end{equation}
	The vectors ${\bf y}_{\bf I}[k]$ form the data for tomographic analysis. 
	
	Our tomographic analysis takes into account the noise properties, including noise correlation. 
	As we now show, the measurement model~[\ref{supeq:y}] yields correlated noise of different Stokes components. Thus, ${\bf R}^{-1}$ (\eqref{eq:minimization}) is not diagonal. Denote the diagonal co-variance matrix of the photo-electron readings by ${\bf C}_{\rm e}^{-1}{=}\text{diag}\big({\bf e}\big)$. Let $\mathcal{I}_{46{\times}46}$ denote the {\em Identity} matrix. The signal is generally dominated by unpolarized multiply-scattered background light. Relative to it, the magnitude of the modulated polarization signal is small. Thus, per pixel $k$, the diagonal matrix ${\bf C}_{\rm e}^{-1}[k]$ is approximately constant with a global weight 
	\begin{equation}
	{\bf C}_{\rm e}^{-1}[k] \approx \alpha[k] \mathcal{I}_{46{\times}46}.
	\label{supeq:C}
	\end{equation}
	Using Eqs.~[\ref{supeq:y},\ref{supeq:C}] for each pixel, the Stokes co-variance matrix is   
	\begin{equation}
	{\bf C}^{-1}[k] = {\bf M}^\top {\bf C}_{\rm e}^{-1}[k] {\bf M} \approx \alpha[k]{\bf M}^\top {\bf M}.
	\label{supeq:pixel_noise}
	\end{equation}
	A maximum-likelihood estimator corresponding to a Poisson process should have a weight $\alpha[k] \propto 1/\|{\bf e}\|_1$, to account for higher photon noise in brighter pixels. In simulations, however, we found that $\alpha[k]=1$ worked better. This is perhaps due to richer information carried by denser cloud regions, i.e. brighter pixels. Overall the expression we minimize in \eqref{eq:minimization} is
	\begin{equation}
	\hat{{\bf \Theta}} = \underset{{\bf \Theta}}{\arg\min}  \sum_{k=1}^{N_{\rm meas}} \left({\bf I}[k] - {\bf y}_{\bf I}[k] \right)^\top {\bf M}^\top {\bf M} \left({\bf I}[k] - {\bf y}_{\bf I}[k] \right),
	\label{supeq:minimization_noise}
	\end{equation}
	i.e. ${\bf R}^{-1} = {\bf M}^\top {\bf M}$.
	

	\section{Numerical considerations}
	\label{subsec:numerical}
	
	In this section we describe numerical considerations that stabilize the recovery.
	
	\subsection{Hyper-parameters}
	Our code requires the choice of hyper-parameters for rendering with vSHDOM~\cite{pincus2009computational} in {\tt Step 1} and optimization with {\em scipy} L-BFGS~\cite{zhu1997algorithm,scipy} in {\tt Step 2}. 
	Table \ref{table:numerical_params} summarizes the numerical parameters used in our simulations.
	\begin{table}[t]
		\centering
		\begin{minipage}{.35\linewidth}
			\centering
			\caption*{vSHDOM}
			\begin{tabular}{ccc} \toprule
				{$N_{\mu}$} & {$N_{\phi}$} & {splitting accuracy}  \\ \midrule
				8 & 16 & 0.1 \\
			\end{tabular}
		\end{minipage}%
		\begin{minipage}{.35\linewidth}
			\centering
			\caption*{L-BFGS}
			\begin{tabular}{ccc} \toprule
				{gtol} & {gtol} & {maxls}  \\ \midrule
				$1{\rm e}^{-16}$ & $1{\rm e}^{-16}$ & 30 \\
			\end{tabular}
		\end{minipage} 
		\vspace{0.2cm}
		\caption{Numerical parameters. For vSHDOM parameter definitions, see Ref.~\cite{pincus2009computational}. For L-BFGS parameter definitions, see Ref.~\cite{scipy}.} 
		\label{table:numerical_params}
	\end{table}
	
	\subsection{Preconditioning}
	Multivariate optimization can suffer from ill-conditioning due to different scales of the sought variables. This is expected when recovered variables represent different physical quantities with different units and orders of magnitude.
	A preconditioning of the update rule in \eqref{eq:update_rule} takes the following form
	\begin{equation}
	{\bf \Theta}_{b+1} = {\bf \Theta}_{b} - \chi_b {\bf \Pi}^{-1} \nabla_{\bf \Theta} \mathcal{D}\left({\bf I}_{\bf \Theta},{\bf y} \right),
	\end{equation}
	where we apply a diagonal scaling matrix ${\bf \Pi}$ ({\em Jacobi} preconditioner) to scale the different physical variables $({\rm LWC}, r_{\rm e})$. Thus, ${\bf \Pi}$ takes the form
	\begin{equation}
	{\bf \Pi} = {\rm diag}\big(\Pi_{\rm LWC},~\Pi_{r_{\rm e}},....,\Pi_{\rm LWC},~\Pi_{r_{\rm e}}\big).
	\end{equation}
	In our tests, we use $\Pi_{\rm LWC}$ = 15 and $\Pi_{r_{\rm e}}$ = 0.01 to scale the parameters to a similar magnitude and closer to unity upon initialization.
	
	\subsection{Initialization}
	The recovery is initialized by the estimation of a cloud voxel mask, which bounds the cloud 3D shape. The 3D shape bound of the cloud is estimated using {\em Space-Carving}~\cite{veikherman2014clouds}. Space-carving is a geometric approach to estimate a bound to 3D shape via multi-view images. The following steps are preformed in our space-carving algorithm
	\begin{enumerate}[leftmargin=0.7cm,itemsep=-1pt,topsep=2pt]
		\item Each image is segmented into {\em potentially cloudy} and {\em non-cloudy} pixels (we use a simple radiance threshold).
		\item From each camera viewpoint, each {\em potentially cloudy} pixel back-projects a ray into the 3D domain. Voxels that this ray crosses are voted as potentially cloudy.
		\item Voxels which accumulate ``cloudy'' votes in at least 8 out of the 9 AirMSPI viewpoints are marked as cloudy. 
	\end{enumerate}
	Outside of the shape bound, LWC = 0 throughout iterations. Within the estimated cloud-shape bound, the volume content is initialized as homogeneous with \mbox{${\rm LWC} = 0.01~\nicefrac{{\rm g}}{{\rm m}^3}$}, \mbox{$r_{\rm e} = 12  \mu{\rm m}$} and $v_{\rm e}$ = 0.1. 
	Then, inside of the shape-bound, \{LWC,$r_{\rm e}$,$v_{\rm e}$\} change throughout iterations, possibly diminishing LWC to very small values.
	
	\subsection{Convergence}
	Our approach alternates between {\tt Step 1} (RTE rendering) and {\tt Step 2} (approximate gradient) until convergence (Fig.~\ref{fig:block_diagram}). The convergence criteria are dictated by the L-BFGS step: at each iteration, the relative change to the forward model and its gradient are compared to the {\em ftol} and {\em gtol} parameters (see Table~\ref{table:numerical_params} for values used). See {\em SciPy} documentation~\cite{scipy} for exact description of the L-BFGS stopping criteria.
	

	\section{Qualitative Results: {\tt Scene B}}
	
\begin{figure}
		\centering 
		\includegraphics[width=0.35\textwidth]{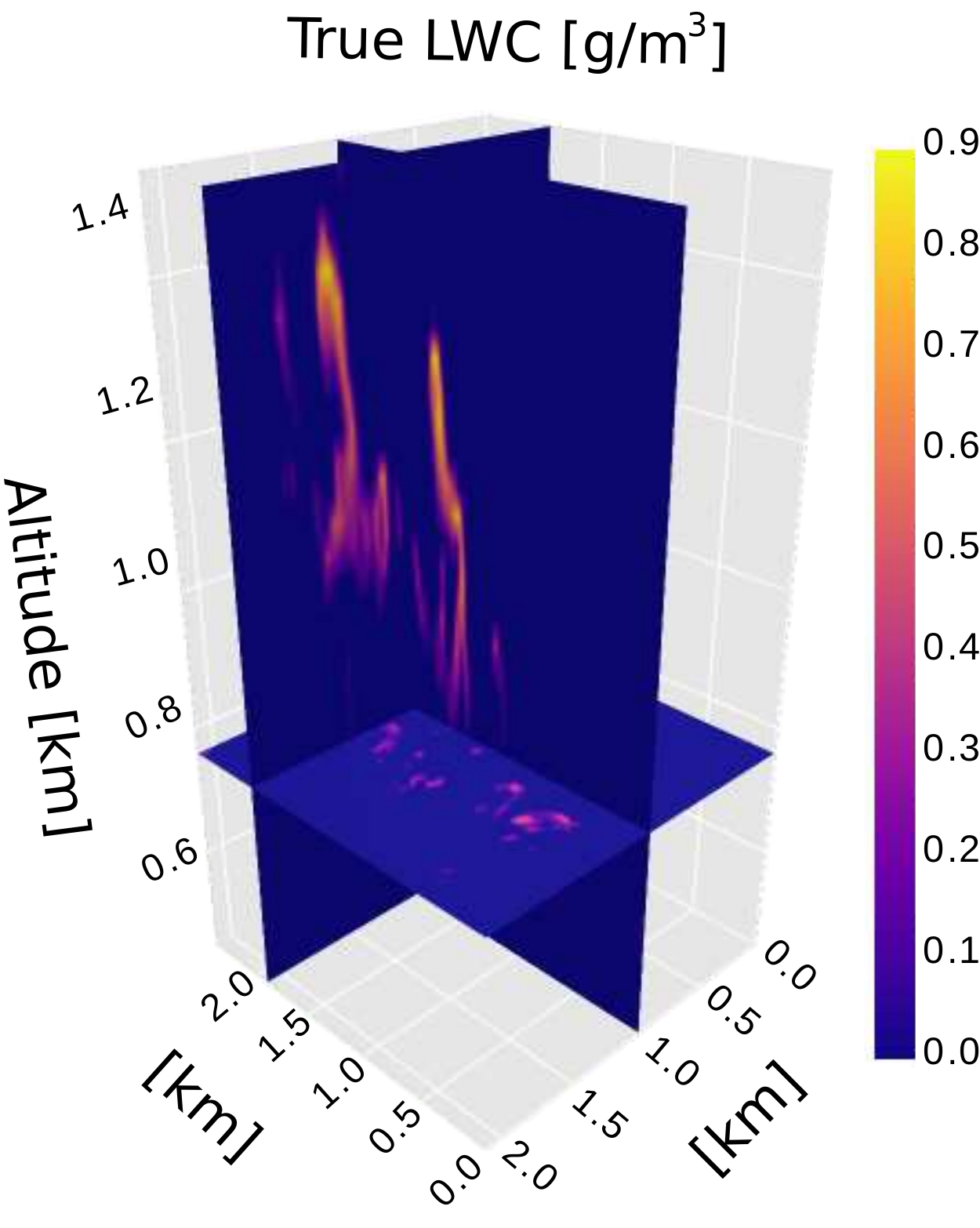} ~~~~~~~~~~~~~
		\includegraphics[width=0.35\textwidth]{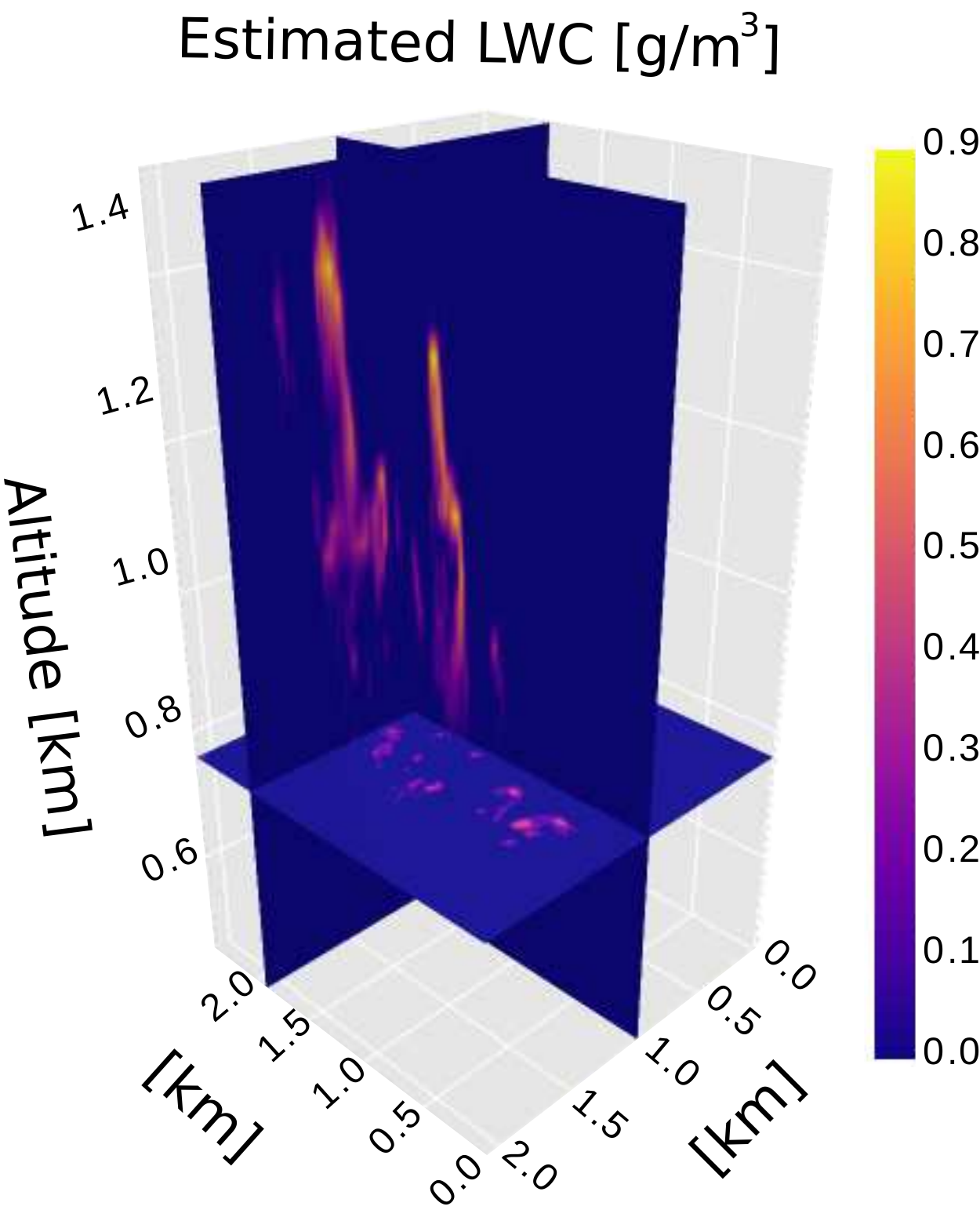}
		\caption{{\tt Scene B} recovery results. [Left] Slices of the true LES generated region. [Right] Slices of the estimated region.}
		\label{fig:lwc2_true}
	\end{figure}

\begin{figure}
		\centering 
		\includegraphics[width=0.35\textwidth]{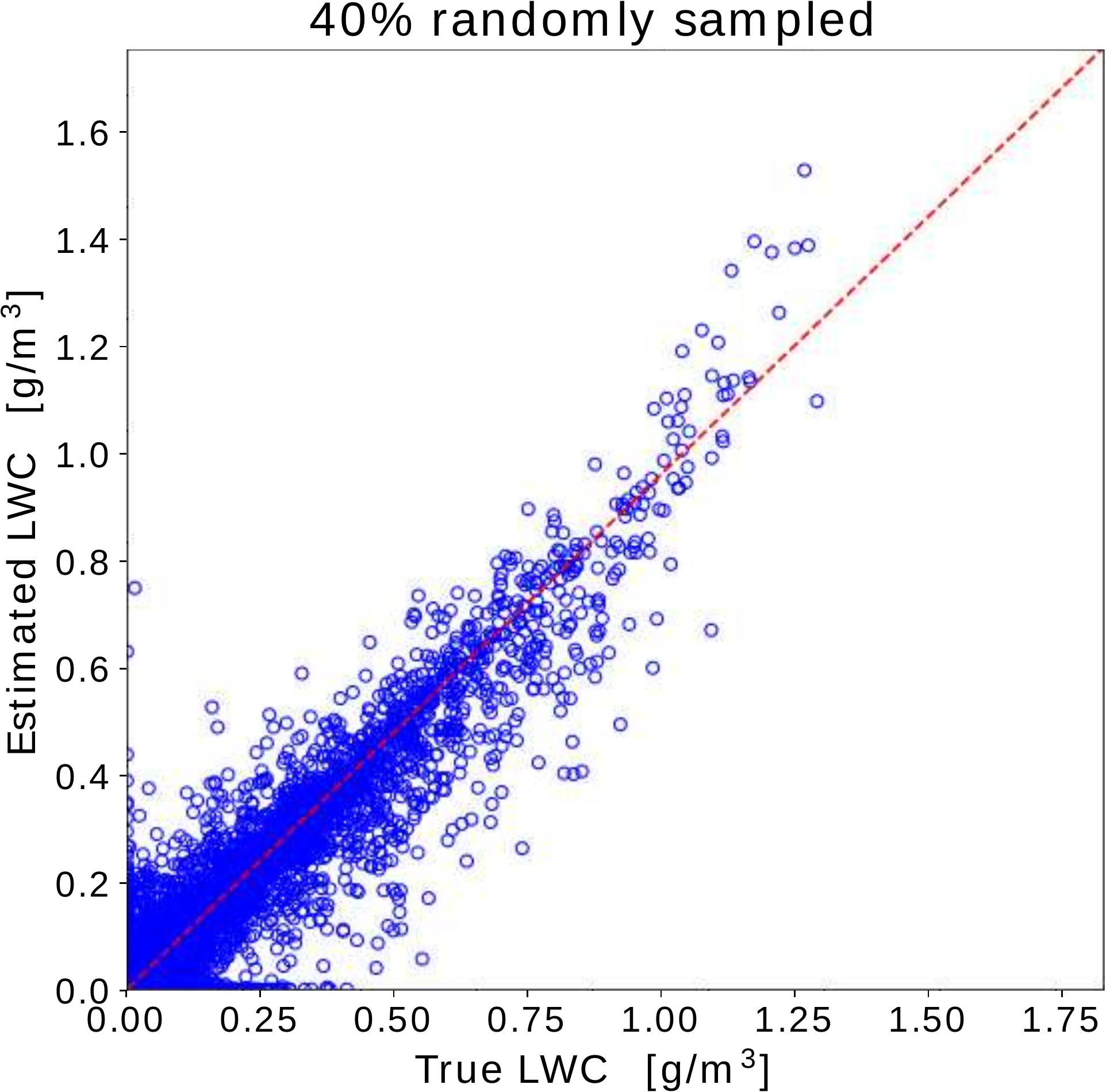}~~~~~~~~~~~~~~~
		\includegraphics[width=0.35\textwidth]{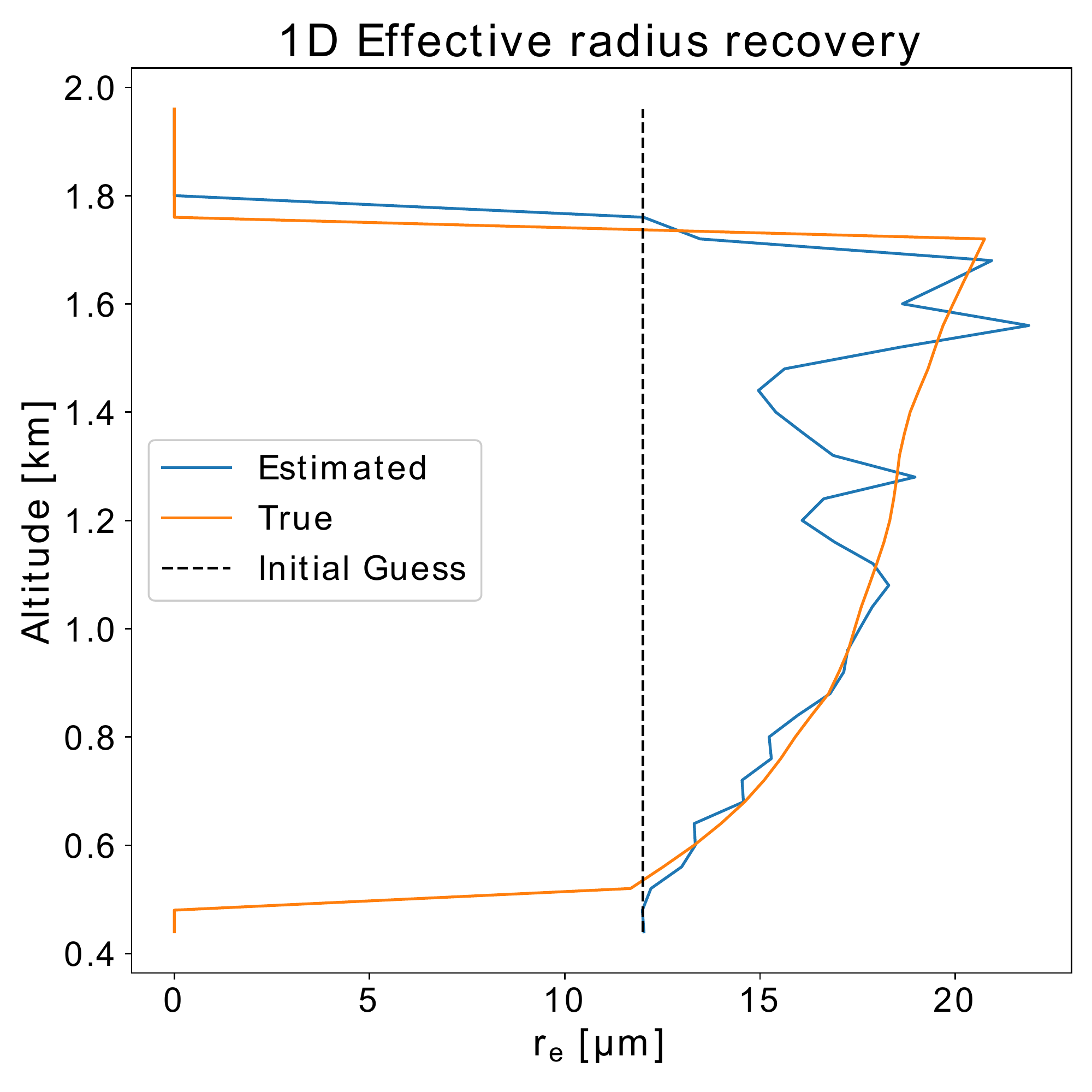}
		\caption{{\tt Scene B} [Left] Scatter plot of the estimated vs. true LWC. The fit correlation is 0.96. [Right] Recovery results of the 1D effective radius}
		\label{fig:scatter2}
	\end{figure}

	Qualitative volumetric results of the recovered LWC for {\tt Scene B} are shown in Fig.~\ref{fig:lwc2_true}. A scatter plot of the recovered LWC and the recovery results of $r_{\rm e}$ for {\tt Scene B} are given in Fig.~\ref{fig:scatter2}.

	\section{Spatial Variation of The Effective Radius}
	\label{sec:discussion}
	
	In nature, generally the droplet effective radius $r_{\rm e}$ and variance  $v_{\rm e}$ vary in 3D. 
	However, operational remote sensing algorithms, which rely on 1D RT and plane-parallel cloud models, retrieve a single value for $r_{\rm e}$ (and for $v_{\rm e}$), for each cloudy pixel.  
	This occurs both in bi-spectral~\cite{nakajima1990determination,EwaldEtAl_19} and polarimetric~\cite{BreonGoloub_98,AlexandrovEtAl_12} techniques. 
	In these approaches, it is always uncertain which portion of the cloud the retrieved quantity corresponds to, because light penetrates into the cloud and simultaneously scatters from different depths inside it. 
	In polarization analysis of plane-parallel cloud models, it is often assumed that the retrieved microphysical parameters correspond approximately to an optical depth of unity. 
	At any rate, this uncertainty complicates the interpretation of retrieved values in studies which rely on them. 
	
	The mathematical approach of the paper is formulated for 3D variation of all the required fields: ${\rm LWC},r_{\rm e},v_{\rm e}$. 
	As Fig.~3 in the main paper shows, polarization is sensitive to $r_{\rm e}$ of any voxel which scatters sunlight towards the camera. Moreover, the formulation explicitly models and seeks spatially varying microphysics, using multi-angular data. 
	We confidently anticipate the same sensitivity to $v_{\rm e}$.
	The demonstrations in the simulations used a representation in which $r_{\rm e}$ varies vertically, not horizontally. 
	This is more general than the operational methods mentioned above, yet more degenerate than full 3D heterogeneity. 
	We now discuss the implication of such a representation. 
	
	
	Textbook cloud physics (e.g.,~\cite{yau1996short}) is based on the mental picture of a parcel of moist air containing a certain number of cloud condensation nuclei that is ascending vertically in the buoyancy-driven part of the convective cycle.  
	Since temperature and pressure are strongly stratified environmental quantities, moist adiabatic thermodynamics thus predict a vertically-varying droplet size distribution, at least in the so-called ``convective core'' of the cloud.
	For the present study, this restriction of microphysical variability to the vertical dimension only applies to both  $r_{\rm e}$ and  $v_{\rm e}$.
	
	There is compelling evidence that the horizontal variability  $r_{\rm e}$ is indeed small over a cloud scale. 
	This evidence comes from in-situ aircraft observations of shallow cumulus~\cite{BlythLatham_91,FrenchEtAl_00,GerberEtAl_08}, modelling studies~\cite{KhainEtAl_19} and theory~\cite{PinskyKhain_18}. 
	However, there are also select observations of monsoonal clouds~\cite{BeraEtAl_16} and theoretical arguments~\cite{PinskyKhain_18} that suggest there is a sharp gradient in the droplet effective radius in the very outer shell of the clouds. 
	If this is the case, then a representation having vertical-only variation of $r_{\rm e}$ loses validity at the outer shell. This may cause bias in retrievals based on polarimetry. 
	The reason is that polarization signals are dominated by single-scattering, which is most likely to occur at shallow depth in the cloud.  
	
	
	The value of $v_{\rm e}$ can also vary significantly across different environmental conditions. 
	This is seen in research flights including in-situ measurements~\cite{CostaEtAl_00,LuEtAl_08,MartinsSilva-Dias_09,HudsonEtAl_12,PandithuraiEtAl_12}. 
	Moreover, in LES simulations of shallow cumulus clouds with bin microphysics, $v_{\rm e}$ might range from 0.01 to 0.26~\cite{Igelvan-den-Heever_17}. 
	The core of a cloud tends to have a low effective variance as condensation is the dominant process there~\cite{LuSeinfeld_06,Igelvan-den-Heever_17}. 
	Cloud edges, in contrast, experience also evaporation and entrainment mixing, as the cloud is diluted by environmental air~\cite{WangEtAl_11}. 
	This tends to increase $v_{\rm e}$. 
	If the cloud has precipitation, spatial variability of $v_{\rm e}$ 
	increases~\cite{MilbrandtYau_05}.
	
	These points show that, on the one hand, the approximations in the demonstrations are often reasonable. 
	On the other hand, it is indeed worth representing cloud microphysical parameters as functions in 3D, then retrieving them in tomography, to push the frontier of cloud physics research. 
	Retrieving a large number of degrees of freedom can be managed better by using more information from diverse sources. 
	One option is to include additional sources of measurements, e.g., by using a combination of the  AirMSPI~\cite{diner2013airborne} and Research Spectro-Polarimeter (RSP)~\cite{cairns1999research} airborne instruments. 
	Another option is to introduce tailored regularization schemes, which mathematically express the natural trends of horizontal variability mentioned above. 
	The 3D tomographic approach presented in the paper is a significant enabler for probing such questions. 
	It offers more flexibility than current operational analyses, which are largely based on 1D RT and bulk retrieved values for a whole cloud.  
	
\end{document}